\RequirePackage[final]{graphicx}
\documentclass[aps, superscriptaddress, prb, twocolumn, 10pt, floatfix, longbibliography]{revtex4-2}
\usepackage[utf8]{inputenc}
\usepackage{amsmath}
\usepackage{amssymb}
\usepackage{amsfonts}
\usepackage{mathtools}
\usepackage{fixme}
\usepackage{color}
\usepackage{braket}
\usepackage{dsfont}
\usepackage{hyperref}

\def\bd{{\bf d}}

\def\bi{{\bf i}}
\def\bj{{\bf j}}

\def\bsigma{{\pmb \sigma}}

\def\calI{\mathcal{I}}

\def\calO{\mathcal{O}}

\def\calP{\mathcal{P}}

\def\nn{\nonumber}

\def\Ham{{ \hat{H} }}

\begin{document}
\title{Beyond fragmented dopant dynamics in quantum spin lattices: \\ Robust localization and non-Gaussian diffusion}
\author{Mingru Yang}
\affiliation{Max Planck Institute of Quantum Optics, Hans-Kopfermann-Str. 1, D-85748 Garching, Germany}
\affiliation{Munich Center for Quantum Science and Technology, Schellingstraße 4, 80799 München, Germany}
\author{Sajant Anand}
\affiliation{Department of Physics, University of California, Berkeley, CA 94720, USA}
\affiliation{Material Science Division, Lawrence Berkeley National Laboratory, Berkeley, California 94720, USA}
\author{Kristian Knakkergaard Nielsen}
\affiliation{Niels Bohr Institute, University of Copenhagen, Jagtvej 128, DK-2200 Copenhagen, Denmark}
\affiliation{Max Planck Institute of Quantum Optics, Hans-Kopfermann-Str. 1, D-85748 Garching, Germany}
\date{\today}

\begin{abstract}
The motion of dopants in magnetic spin lattices has received tremendous attention for at least four decades due to its connection to high-temperature superconductivity. Despite these efforts, we lack a complete understanding of their behavior, especially out of the equilibrium and at nonzero temperatures. In this paper, we take a significant step towards a much deeper understanding based on state-of-the-art matrix-product-state calculations. In particular, we investigate the non-equilibrium dynamics of a dopant in two-leg $t$--$J$ ladders with antiferromagnetic XXZ spin interactions. In the Ising limit, we find that the dopant is \emph{localized} for all investigated \emph{nonzero} temperatures due to an emergent disordered potential, with a localization length controlled by the underlying correlation length of the spin lattice, which increases exponentially with decreasing temperature. The dopant, hereby, only delocalizes asymptotically in the zero temperature limit. This greatly generalizes the localization effect discovered recently in Hilbert space fragmented models [PRResearch \textbf{6}, 023325 (2024), SciPost Phys. Core \textbf{7}, 054 (2024)]. In the presence of spin-exchange processes at rate $\alpha$, the dopant diffuses with a diffusion coefficient, $D_h$, depending non-monotonically on $\alpha$. It initially increases linearly as $D_h \propto \alpha$ for $\alpha \ll 1$ before dropping off as $\alpha^{-1}$ for $\alpha > 1$. Moreover, we show that the underlying spin dynamics at infinite temperature behaves qualitatively the same, albeit with important quantitative differences. We substantiate these findings by showing that the dynamics features self-similar scaling behavior, which strongly deviates from the Gaussian behavior of regular diffusion, especially for weak spin exchange. Finally, we show that the diffusion coefficient $D_h$ follows an Arrhenius relation at high temperatures, whereby it is exponentially suppressed upon cooling. 
\end{abstract}

\maketitle

\section{Introduction}
Tremendous effort has been dedicated to the study of emergent many-body properties at low energy and in equilibrium. While generic quantum many-body systems are exponentially complex in system size, static and local interactions at low energies typically lead to states with only area-law entanglement \cite{Hastings2007,Eisert2010}. This emergent structure has enabled the success of tensor network algorithms \cite{White1992}. In recent years, however, quantum simulators have pushed beyond this regime and, e.g., investigated dynamical phase transitions \cite{Jurcevic2017,Karch2025}, quantum many-body scars \cite{Bernien2017,Moudgalya2022} and Hilbert space fragmentation \cite{Moudgalya2022b,Adler2024}. This has unveiled that emergent structures are not limited to low energies and equilibrium; they may appear dynamically and throughout the many-body spectrum as well. 

For ultracold atoms in optical lattices in particular, the awesome capabilities of single-site resolved quantum gas microscopy \cite{Bakr2009,Sherson2010,Cheuk2015,Haller2015,Gross2021,Wei_2022} call for more precise theoretical descriptions of the rich behaviors in the dynamics of local observables. More specifically, such systems naturally implement Fermi-Hubbard and $t$--$J$ models \cite{Gross2017,Tarruell2018,Schafer2020} and allow us to probe the intricate interplay of charge and spin degrees of freedom \cite{Greif2013,Hart2015,Boll2016,Parsons2016,Cheuk2016b,Mazurenko2017,Mitra2018,Hartke2023}, which is at the core of our understanding of strongly correlated electronic systems, including high-temperature superconductors \cite{highTc}. 

A central interest in this respect is to study with great detail the microscopic behavior of charges (dopants) both in \cite{Koepsell2019,Chalopin2024} and out of equilibrium \cite{Ji2021}. It is generally believed that dopant motion leads to the formation of itinerant quasiparticles \cite{SchmittRink1988,Kane1989}, which is also consistent with experiments so far \cite{Nielsen2022_2}. There is currently, however, 
still no explicit evidence for their existence 
\cite{Nielsen2025_2}. This is echoed on the theoretical side, where we have so far lacked high-accuracy results. For example, it is not clear whether quasiparticles are present in general $t$--$J$ models \cite{Sheng1996,Wu2008,Zhu2015,White2015,Sun2019,Zhao2022}, although  matrix product state (MPS) calculations have shown robust magnetic polaron quasiparticles in ladders \cite{White2015} and quasi one-dimensional (1D) cylindric geometries at \emph{zero} temperature \cite{Bohrdt2020}, leading to ballistic transport of dopants. At nonzero temperatures, thermal damping is expected to break this quasiparticle behavior. Indeed, initial studies at \emph{infinite} temperatures in two-dimensional (2D) square lattices revealed slower than ballistic behavior \cite{Carlstrom2016,Nagy2017}, though the actual long-time behavior remains illusive.

\begin{figure}[t!]
    \centering
    \includegraphics[width=0.8\columnwidth]{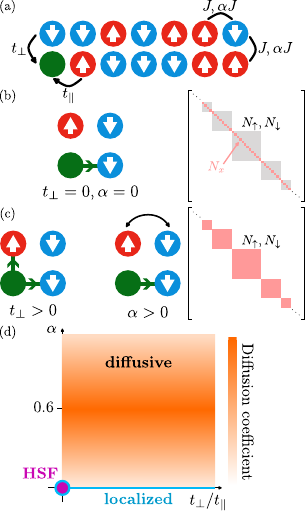}
    \vspace{-0.25cm}
    \caption{\textbf{Setup and breaking Hilbert space fragmentation.} (a) The system consists of two spin states (red: $\uparrow$, blue: $\downarrow$) and a single dopant -- a hole -- shown in green. The spins interact via Ising interactions $J$ and spin exchange $\alpha J$ and may hop onto the vacant site with amplitudes depending on the direction of the hop ($t_\parallel,t_\perp$). (b) The $t$--$J_z$ limit, i.e. Ising-type interactions ($\alpha = 0$) with only one-dimensional hopping ($t_\perp = 0$), features Hilbert space fragmentation. For a single hole, each block of fixed magnetization (fixed $N_\uparrow,N_\downarrow$) split into Krylov subspaces that are only the length of the ladder, $N_x$, large. (c) For any nonzero rung hopping ($t_\perp > 0$) \emph{or} any spin exchange ($\alpha > 0$), each block with fixed magnetization collapses to a single Krylov subspace specified by the $U(1)$ spin conservation. (d) High-temperature dynamical phase diagram: while the fragmented phase (HSF) is special to $\alpha = t_\perp = 0$, the dopant remains localized for any nonzero $t_\perp$ and $\alpha = 0$ (blue). For any $\alpha > 0$, the dopant delocalizes diffusively. Moreover, the associated diffusion coefficient is maximal at intermediate spin exchange around $\alpha = 0.6$ as schematically indicated by the orange color gradient. }
    \vspace{-0.25cm}
    \label{fig:HSF}
\end{figure}

Understanding the temperature dependence of dopant dynamics, therefore, remains an outstanding and fundamental challenge. In this work, we address this issue by investigating a minimal setup of two-leg $t$--$J$ ladders with antiferromagnetic XXZ spin interactions [Fig.~\ref{fig:HSF}(a)]. We generalize a previously studied Hilbert space fragmented model [Fig.~\ref{fig:HSF}(b)], in which dopants are localized at any nonzero temperature and spin interaction \cite{Nielsen2023_3}. In the Ising limit, we show that localization is robust, even though fragmentation breaks down, and find that the localization length scales with the underlying spin-spin correlation length. In the presence of spin-exchange processes, we find that both dopant and spins diffuse, signaled by  a mean-square distance growing linearly in time. Moreover, in this diffusive regime we find that the transport features self-similar scaling behavior, which strongly deviates from the standard Gaussian behavior seen in regular diffusion \cite{Crank_1975}, indicating that the dynamics is hard to describe by simple hydrodynamic relations. Finally, we show that the diffusion coefficient \emph{decreases} exponentially according to an Arrhenius relation \cite{Arrhenius1889} with decreasing temperatures. \\

The rest of the Article is organized as follows. In Sec. \ref{sec.fragmentation_and_beyond}, we review the previously described fragmented models \cite{Nielsen2023_3,Nielsen2024_1} and describe our main results for how transport is affected once we leave this regime. In Sec. \ref{sec.methods}, we give a brief description of the numerical methods used. In Sec. \ref{sec.ising_limit}, we investigate the impact of nonzero rung hoppping in the Ising limit. In Sec. \ref{sec.spin_exchange}, we investigate the impact of spin-exchange processes. Finally, in Sec. \ref{sec.towards_2D} we speculate on the behavior in fully 2D spin lattices, before we conclude in Sec. \ref{sec.conclusions}.

\section{Fragmented \texorpdfstring{$t$-$J_z$}{t-Jz} models and beyond} \label{sec.fragmentation_and_beyond}
Systems with dopants moving \emph{only in 1D} via nearest-neighbor hopping and spins constrained to Ising-type interactions feature Hilbert space fragmentation \cite{Moudgalya2022b}, in which a sector with fixed numbers $N_\uparrow,N_\downarrow$ of spin-$\uparrow,\downarrow$ shatters into exponentially many Krylov subspaces as a function of system size [Fig.~\ref{fig:HSF}(b)]. Indeed, taking a two-leg ladder as an example with an initial configuration of one hole, $\ket{0}$, at one end, 
\begin{equation}
\ket{\psi_0} = \ket{\substack{\uparrow\;\uparrow\;\downarrow\;\uparrow\;\uparrow\;\downarrow \;\downarrow\;\uparrow \,\dots\\
0\;\downarrow\;\uparrow\;\downarrow\;\uparrow\;\uparrow\;\downarrow \;\uparrow \,\dots}}\phantom{,\dots}
\end{equation}
only couples to states of the form
\begin{align}
\ket{\psi_1} &= \ket{\substack{\uparrow\;\uparrow\;\downarrow\;\uparrow\;\uparrow\;\downarrow \;\downarrow\;\uparrow \,\dots\\
\downarrow\;0\;\uparrow\;\downarrow\;\uparrow\;\uparrow\;\downarrow \;\uparrow \,\dots }}, \nn \\
\ket{\psi_2} &= \ket{\substack{\uparrow\;\uparrow\;\downarrow\;\uparrow\;\uparrow\;\downarrow \;\downarrow\;\uparrow \,\dots\\
\downarrow\;\uparrow\;0\;\downarrow\;\uparrow\;\uparrow\;\downarrow \;\uparrow \,\dots}}, \dots 
\end{align}
in which the hole only moves along the ladder (the hopping amplitude across the ladder is here set to $0$). Consequently, the underlying spin patterns in each leg of the ladder, $\ket{\uparrow\;\uparrow\;\downarrow\;\uparrow\;\uparrow\;\downarrow \;\downarrow\;\uparrow \,\dots}$ and $\ket{\downarrow\;\uparrow\;\downarrow\;\uparrow\;\uparrow\;\downarrow \;\uparrow \,\dots}$, remain unaltered and constitute the distinct Krylov subspaces. Importantly, this feature immediately leads to fragmentation and is unaltered by the inclusion of more legs in the lattice as well as an arbitrary number of dopants as long as the dopant only moves along one direction in the lattice. This can be seen by generalizing the arguments made in Ref. \cite{Moudgalya2022b}, on which we elaborate in Appendix \ref{app:fragmentation}. In turn, the unchanging spin pattern enables the study of thermodynamically large systems, revealing that the dopants are localized for \emph{any} nonzero spin coupling, both ferro- and antiferromagnetic, and at \emph{any} nonzero temperature \cite{Nielsen2023_3}. This remains true even when the system crosses the Ising phase transition to a ferromagnetic phase in a 2D square lattice \cite{Nielsen2024_1} by tuning the temperature. Moreover, the fragmented model features strong non-equilibrium \emph{pairing} of dopants at infinite temperatures \cite{Nielsen2025_1}. The underlying localization and pairing mechanism was traced back to an \emph{emergent} disordered potential for the dopants. In more novel terms, there is a direct link between this localization effect and \emph{disorder-free localization} in lattice gauge theories with static charges \cite{Smith2017,Gyawali2025}. Indeed, both systems can be mapped to an Anderson model of free fermions with onsite disorder \cite{Anderson1958}, utilizing the underlying Hilbert space fragmentation of the respective models \footnote{In disorder-free localization, the local gauge symmetries commute and are responsible for the Hilbert space fragmentation.}. A natural question to ask, therefore, is: \\

\emph{Is the localization mechanism tied to Hilbert space fragmentation?} \\

\noindent That is, does localization only happen in the fragmented models, or is it possible to retain localization in more generic situations? Moreover, what kind of transport behavior do we expect if localization breaks down when the Hilbert space is no longer fragmented? Will the dopants, e.g., immediately start to diffuse?

To simplify the problem but still retain the essential 2D physics, we focus our studies on a minimal two-leg ladder geometry of system size $N=N_x\times 2$ subject to the $t$--$J$ Hamiltonian 
\begin{align}
\!\! \Ham = \Ham_t \!+ \Ham_J{} ={}& -\!\!\sum_{\substack{\lambda =\parallel,\perp\\\sigma = \uparrow,\downarrow}}\!\sum_{\braket{\bi,\bj}_\lambda} t_\lambda \hat{\cal{P}}\left[\hat{c}^\dagger_{\bi\sigma}\hat{c}_{\bj\sigma} \!+\! \hat{c}^\dagger_{\bj\sigma} \hat{c}_{\bi\sigma}\right]\hat{\cal{P}}\nn \\
&+ J \sum_{\braket{\bi,\bj}} \left[\hat{S}_\bi^{z}\hat{S}_\bj^{z} \!+\! \alpha\left(\hat{S}_\bi^{x}\hat{S}_\bj^{x} \!+\! \hat{S}_\bi^{y}\hat{S}_\bj^{y}\right)\right].\!\!
\end{align} 
This describes nearest-neighbor (NN) constrained hoppings $t_\parallel, t_\perp$, of spin-$1/2$ fermions interacting via antiferromagnetic XXZ-type NN spin interactions $J>0$ [Fig.~\ref{fig:HSF}(a)]. Indeed, $\hat{c}^\dagger_{\bi\sigma}$ ($\hat{c}_{\bi\sigma}$) creates (destroys) a fermion at site $\bi$ with spin-$z$ $\sigma = \uparrow,\downarrow$, while $\hat{\cal{P}}$ projects onto the subspace of at most one fermion per site. In experiments, such models may naturally be implemented by combining Fermi-Hubbard simulators using ultracold atoms in optical lattices \cite{Greif2013,Hart2015,Boll2016,Parsons2016,Cheuk2016b,Mazurenko2017,Mitra2018,Hartke2023, Koepsell2019,Ji2021,Chalopin2024} with Rydberg dressing \cite{Zeiher2017,Weckesser2024}, or alternatively using dipole-dipole and van der Waals interactions in Rydberg arrays \cite{Homeier2024,Qiao2025}. Our findings are, therefore, not only fundamentally important for understanding these systems, but may also be tested in state-of-the-art quantum simulators.  

The aforementioned fragmented $t$--$J_z$ limit is recovered when both the rung hopping and the spin exchange vanish, $t_\perp = \alpha = 0$. We then consider two \emph{distinct} ways of breaking fragmentation: (1) a nonzero rung hopping, $t_\perp > 0$, and (2) a nonzero spin exchange, $\alpha > 0$. In the latter case, the ensuing exchange processes connect all spin configurations with the same conserved quantum numbers, $N_\uparrow,N_\downarrow$. In the former case with Ising interactions, nonzero hopping along both spatial directions has the same effect and is rooted in the possibility of moving in loops in the lattice \cite{Carlstrom2016,Bobrow2018}. In turn, both scenarios ($t_\perp > 0$ and $\alpha > 0$) break Hilbert space fragmentation, in that the many Krylov sub-sectors for given $N_\uparrow,N_\downarrow$ now implode to one single $U(1)$ block [Fig.~\ref{fig:HSF}(c)].

One might expect that localization breaks down in both cases, because the system is now in principle \emph{ergodic}. However, we show that the two scenarios play out in drastically distinct ways. For the $t$--$J_z$ limit with nonzero \emph{rung hopping} $t_\perp$, we show that localization remains at all investigated \emph{nonzero} temperatures and that the basic mechanism in terms of an emergent disordered energy landscape is upheld. However, for nonzero spin exchange processes $\propto \alpha J$, the ensuing transport of \emph{spins} immediately leads to delocalization of dopants. Furthermore, we give strong numerical evidence that both dopant and spin transport are diffusive for any nonzero $\alpha$. More quantitatively, we find that for both spins and dopant, the diffusion coefficient initially increases linearly with $\alpha$. For larger $\alpha$, the diffusion coefficient for the spins continues to grow monotonically with $\alpha$, whereas for the dopant it peaks around $\alpha = 0.6$ and decreases $
\sim\alpha^{-1}$ for $\alpha > 1$. These results are summarized in the high-temperature dynamical phase diagram in Fig.~\ref{fig:HSF}(d) and imply that the dopant only thermalizes with its spin environment away from the Ising limit of $\alpha = 0$. Finally, we show that in the diffusive regime, both dopant and spin dynamics feature self-similar scaling, which strongly deviates from Gaussianity otherwise inherent to normal diffusion \cite{Crank_1975}. 

\section{Setup and methods} \label{sec.methods}
To track the non-equilibrium dynamics of a single dopant at temperature $k_BT=1/\beta$, we initialize the system with one spin per site (unit filling) in a thermal Gibbs state of the spins, $\hat{\rho}_J = \exp(-\beta \Ham_J)/Z_J$ with $Z_J = {\rm tr}[\exp(-\beta \Ham_J)]$ being the spin partition function of the canonical ensemble. We then create a single hole at the origin (bottom site of left-most rung), 
\begin{equation} \label{eq:initial_state}
\hat{\rho}(0) = \sum_{\sigma_{\bf 0}} \hat{c}_{{\bf 0},\sigma_{\bf 0}} \hat{\rho}_J \hat{c}_{{\bf 0},\sigma_{\bf 0}}^\dagger,
\end{equation}
and time evolve it. In order to calculate the expectation value of some time-evolved observable, $\hat{A}(\tau) = \exp(+i\Ham\tau) \hat{A} \exp(-i\Ham\tau)$, we sample the thermal ensemble through the minimally entangled typical thermal state (METTS) \cite{White2009} approach, i.e.
\begin{align}
\label{eq.monte_carlo_sampling_dynamics}
\braket{\hat{A}(\tau)} &= {\rm tr}[\hat{A}(\tau) \hat{\rho}(0)] = \sum_{\sigma_{\bf 0}} {\rm tr} [e^{+i\Ham\tau} \hat{A} e^{-i\Ham\tau}\hat{c}_{{\bf 0},\sigma_{\bf 0}} \hat{\rho}_J \hat{c}_{{\bf 0},\sigma_{\bf 0}}^\dagger] \nn \\
&= \frac{1}{Z_J}\sum_{\sigma_{\bf 0}} \sum_{\bsigma} p(\bsigma) \bra{\phi_{\sigma_{\bf 0}}(\bsigma,\tau)} \hat{A} \ket{\phi_{\sigma_{\bf 0}}(\bsigma,\tau)}.
\end{align}
Consequently, $\braket{\hat{A}(\tau)}$ is computed as the statistical average over the sample states
\begin{equation} \label{eq.sampled_pure_state}
\ket{\phi_{\sigma_{\bf 0}}(\bsigma,\tau)} = e^{-i\hat{H}\tau} \hat{c}_{{\bf 0},\sigma_{\bf 0}} \ket{\phi(\bsigma)}
\end{equation}
according to the probability distribution $p(\bsigma)/Z_J$, with
$p(\bsigma) = \bra{\bsigma} \exp(-\beta \hat{H}_J)\ket{\bsigma}$ the unnormalized probability for the initial thermal state to be in the product spin state $\ket{\bsigma} = \otimes_\bi \ket{\sigma_\bi}$. Note that we use $\tau$ as the time variable to distinguish it from the hopping amplitude $t_\parallel,t_\perp$. To prepare $\ket{\phi_{\sigma_{\bf 0}}(\bsigma,\tau)}$, we, thus, first imaginary-time evolve a sampled product state to obtain a METTS, i.e. $\ket{\phi(\bsigma)} = p(\bsigma)^{-1/2} \exp(-\beta\Ham_J/2)\ket{\bsigma}$, then remove a spin $\sigma_{\bf 0}$ from the origin, ${\bf 0}$, and finally real-time evolve it to time $\tau$. The imaginary and real time evolution are performed using the time-dependent variational principle (TDVP)~\cite{Yang2020,Haegeman2016,Haegeman2011} and the time evolving block decimation (TEBD) \cite{Vidal2003,White2004}, respectively. For TEBD, we use SWAP gates to switch the interacting sites in the two-leg ladder geometry to nearest neighbor in MPS \cite{White2015}. See Appendix \ref{app:TEBD_details} for more details. We note that in each sampled trajectory, Eq.~\eqref{eq.sampled_pure_state}, we utilize that both the total particle number, $N_\uparrow + N_\downarrow - 1$, and the total magnetization, $(N_\uparrow - N_\downarrow)/2$, are conserved quantum numbers (Appendix \ref{app:METTS_details}). To achieve \emph{high-accuracy} and \emph{unbiased} results, we generally fix the maximum truncation error in each time step and let the bond dimension grow freely. 
The product states $\ket{\bsigma}$ in METTS can be sampled by generating a Markov chain through an efficient state-collapse step \cite{Stoudenmire2010}. In the Ising limit, $\alpha = 0$, one can further simplify the sampling, because now $p(\bsigma) = \exp(-\beta E_J(\bsigma))$, where $E_J(\bsigma)$ is the magnetic energy of the Ising state $\ket{\bsigma}$. Thus, one can use regular Metropolis-Hasting algorithms \cite{Metropolis1953,Hastings1970} for the sampling in this limit.

The bulk of the analysis focuses on the dynamics of the hole density, setting $\hat{A} = \hat{n}_h(\bi) = 1 - \sum_{\sigma} \hat{c}_{\bi\sigma}^\dagger \hat{c}_{\bi\sigma}$, and the main observable is the root-mean-square (rms) distance of the hole to its origin, $\bi_0 = {\bf 0}$ (included here for conceptual clarity),
\begin{align} \label{eq.root_mean_square_distance}
d_{\rm rms}(\tau) = \left[\sum_{\bi} (\bi-\bi_0)^2 n_h(\bi,\tau) \right]^{1/2},
\end{align}
with $n_h(\bi,\tau) = \braket{\hat{n}_h(\bi,\tau)}$. Therefore, localization  corresponds to a bounded rms distance, $d_{\rm rms}(\tau) \leq C$ for all $\tau$, whereas ballistic and diffusive motion manifests as $d_{\rm rms}(\tau) \propto \tau$ and $d_{\rm rms}(\tau) \propto \tau^{1/2}$, respectively, for $\tau \to \infty$. We choose to initialize the dopant at the left end of the two-leg ladder to minimize the required system sizes. While this certainly results in boundary effects at short times, it does \emph{not} alter the nature of the long-time propagation, i.e. whether it is ballistic, diffusive, or localized. We also note that at sufficiently short times, $t_\parallel \tau, t_\perp \tau\ll 1$, the dopant \emph{always} moves independently of the spin background with the ballistic speed of a free particle with nearest neighbor hopping amplitudes $t_\parallel,t_\perp$ \cite{Nielsen2022_2}, such that $d_{\rm rms}(\tau) = [t_\parallel^2 + t_\perp^2]^{1/2} \times \tau$ in this regime. 

\begin{figure*}[t!]
    \centering
    \includegraphics[width=1.0\textwidth]{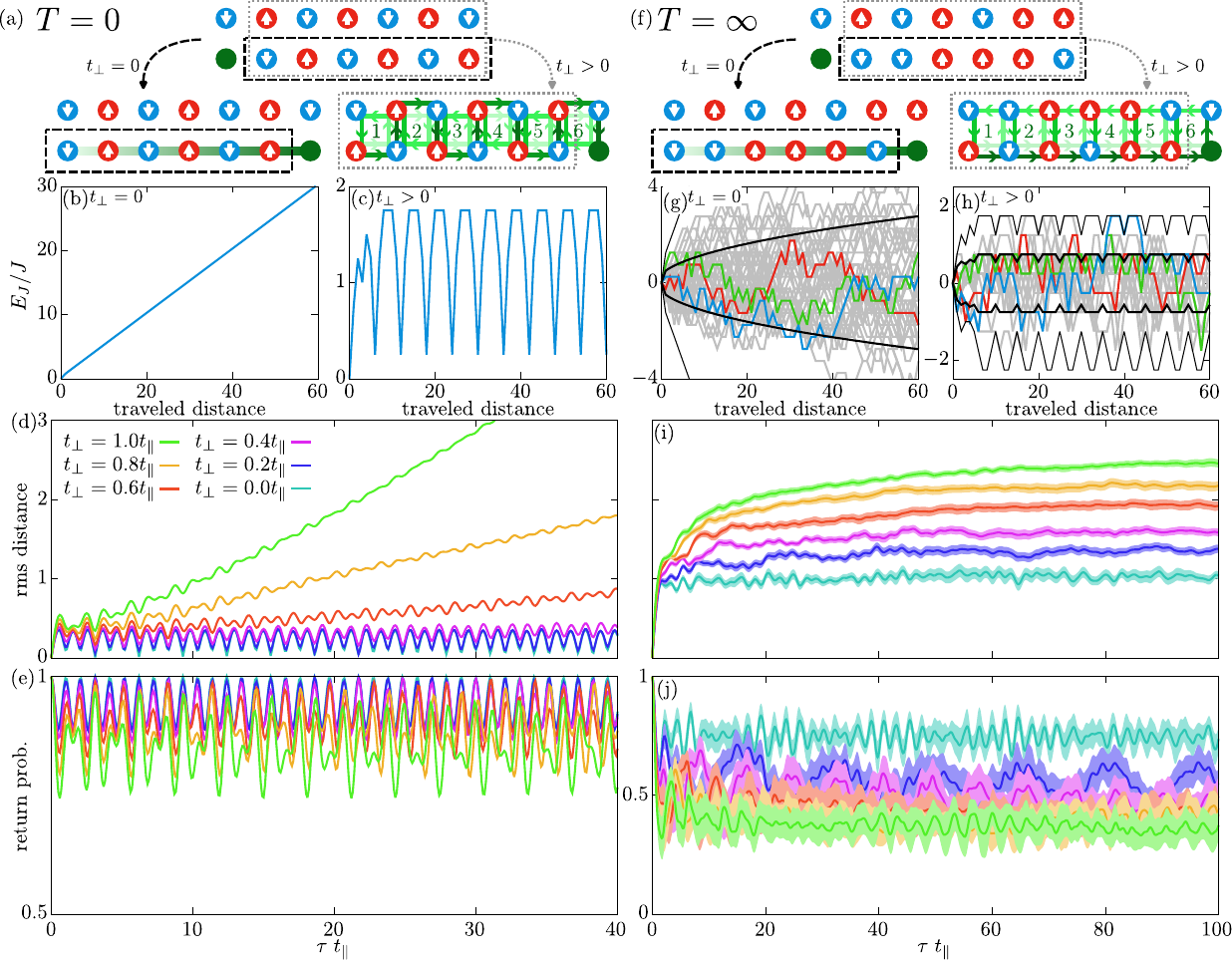}
    \vspace{-0.5cm}
    \caption{\textbf{Dopant dynamics in the Ising limit.} Dopant dynamics at zero temperature (a)-(e), and infinite temperature (f)-(j) in the Ising limit ($\alpha=0$) for indicated values of the rung hopping and $J = 8t_\parallel$. (a) and (f) For $t_\perp = 0$, the spins can only move one site to the left [black long-dashed box]. $t_\perp > 0$ enables Trugman loops shown by the light to dark green arrows around numbered (1-6) $2\times 2$ plaquettes. This moves the entire spin pattern one rung to the left and flips it on its head. [(b) and (c)]  Magnetic energy experienced along the specified paths in the zero temperature N{\'e}el ground state. (g),(h) Magnetic energy along the specified paths at \emph{infinite temperature} shown for $50$ spin realizations (gray lines), with three highlighted in color (red, blue and green lines). Thick black lines show the standard deviation, ${\rm std}(E_J)$, averaged over all spin configurations. Thin black lines show the minimal and maximal values of $E_J$. [(d) and (i)] Resulting rms distance dynamics of the dopant featuring long-time ballistic behavior at zero temperature and robust localization at infinite temperature as the rung hopping is turned on. [(e) and (j)] In both cases, the probability of the dopant returning to the origin remains nonzero. Shaded areas show the estimated standard errors on the mean from the $200$ samples of the spin background. In the TEBD calculations, we use a maximum truncation error of $10^{-8}$ and a system sizes of $N_x \times N_y = 61\times 2$ ($T = 0$) and $N_x \times N_y = 25\times 2$ ($T = \infty$).}
    \vspace{-0.25cm}
    \label{fig:zero_vs_inf_temp_tperp}
\end{figure*}

\section{Ising limit} \label{sec.ising_limit}
We start our analysis in the Ising limit and investigate the impact of turning on the rung hopping $t_\perp$. To this end, we compare the ensuing dynamics of placing the dopant at the left end of the ladder in the zero temperature N{\'e}el ground state to the infinite temperature behavior of completely randomized spins. \\ 

For $t_\perp = 0$, the dopant can only move the spin pattern in the bottom leg one site to the left as shown by the black long-dashed box in Figs. \ref{fig:zero_vs_inf_temp_tperp}(a) and \ref{fig:zero_vs_inf_temp_tperp}(f). At zero temperature, dopant motion in the N{\'e}el ground state, $\ket{\psi_{
\rm N}}$, leads to $l$ pairs of spins that go from being antialigned with energy $-J/4$ to aligned with energy $+J-4$ after a traveled distance of $l$ along each passed rung (see Fig.~\ref{fig:zero_vs_inf_temp_tperp}(b)). As a result, the total magnetic energy change is $E_J(l) = \bra{\psi_{
\rm N}(l)}\Ham_J\ket{\psi_{
\rm N}(l)} - \bra{\psi_{
\rm N}(0)}\Ham_J\ket{\psi_{
\rm N}(0)}$ $ = [J/4 - (-J/4)] \times l = J/2 \times l$ \cite{Nielsen2023_1}. Here, $\ket{\psi_{
\rm N}(l)}$ denotes the state after the dopant has hopped $l$ times along the straight path above Fig.~\ref{fig:zero_vs_inf_temp_tperp}(b). Any $t_\perp > 0$ enables Trugman loops \cite{Trugman1988}. Indeed, when the dopant makes \emph{6 hops} counter-clockwise around a $2\times 2$ plaquette, all spins are moved clockwise by 2 sites. Performing several of such loops consecutively [green arrows around numbered plaquettes in  Fig.~\ref{fig:zero_vs_inf_temp_tperp}(a)] moves the entire staggered magnetization pattern one rung to the left and flips it upside down. For a N{\'e}el ground state, the associated magnetic energy $E_J(l)$ along such a path is periodic with a periodicity of $6$. As a result, the dopant delocalizes, as shown in Fig.~\ref{fig:zero_vs_inf_temp_tperp}(d), where the rms distance of the dopant to its initial position eventually grows linearly at long times as the rung hopping is increased. We note, however, that the long-time propagation speed strongly depends on $t_\perp$, such that the data for $t_\perp = 0.2,0.4$ still shows only very small propagation distances on the plotted timescales. Interestingly, the probability of the dopant to return to its origin, $n_h(\bi = {\bf 0},\tau)$, in Fig.~\ref{fig:zero_vs_inf_temp_tperp}(e) remains high, showing that the dopant density profile has a localized core and delocalized wings.

At infinite spin temperatures, the dopant moving in a lattice with $t_\perp = 0$ leads to a \emph{disordered} magnetic energy landscape, as shown in Fig.~\ref{fig:zero_vs_inf_temp_tperp}(g). Here, we plot the magnetic energy $E_J(\bsigma,l) = \bra{\psi_\bsigma(l)} \Ham_J\ket{\psi_\bsigma(l)} - \bra{\psi_\bsigma(0)} \Ham_J\ket{\psi_\bsigma(0)}$ after the dopant has moved $l$ sites along the straight path shown above Fig.~\ref{fig:zero_vs_inf_temp_tperp}(g) for $50$ randomly chosen spin configurations $\bsigma$. This disordered magnetic potential emerges because when the dopant moves along a straight path, the spins on a specific rung randomly reconfigures from being aligned to anti-aligned or vice versa \cite{Nielsen2024_1}. This leads to the disorder-induced Anderson localization \cite{Anderson1958}, even though there is no quenched disorder in the Hamiltonian. As such, this phenomenon is very similar to the disorder-free localization effect discovered for lattice gauge theories with static gauge fields \cite{Smith2017,Gyawali2025}. One key difference is that the disorder strength here \emph{increases with distance traveled}. Indeed, while the emergent magnetic potential $E_J(\bsigma,l)$ experienced by the dopant on average vanishes across the spin samples, $\mu[E_J(\bsigma,l)] = 0$, the standard deviation from sample to sample, ${\rm std}[E_J(\bsigma,l)] = ({\rm Var}[E_J(\bsigma,l)])^{1/2}\propto |J| \times l^{1/2}$, grows as a square root with distance $l$ \cite{Nielsen2023_3,Nielsen2024_1}. This is shown in solid black lines in Fig.~\ref{fig:zero_vs_inf_temp_tperp}(g). 

Trugman loops still move the entire spin ladder one rung to the left and flip it upside down, as shown above Fig.~\ref{fig:zero_vs_inf_temp_tperp}(h). Crucially, however, the magnetic energy landscape, $E_J(\bsigma,l)$, experienced by the dopant along such a path \emph{remains} disordered, although the disorder strength now becomes \emph{bounded} as a function of distance [Fig.~\ref{fig:zero_vs_inf_temp_tperp}(h)]. As a result, even though the system is no longer fragmented, the dopant remains localized for any rung hopping, as shown by the finite rms distance in Fig.~\ref{fig:zero_vs_inf_temp_tperp}(i) and in stark contrast to the zero-temperature behavior. The dopant is, nevertheless, seen to spread out more for increasing $t_\perp$. Moreover, the dynamics also has a slower approach to its final steady state. As a direct result of the localization, the return probability in Fig.~\ref{fig:zero_vs_inf_temp_tperp}(j) remains nonzero. In turn, this localization effect implies that the dopant retains a memory of its initial state and cannot thermalize with its spin environment. 

\subsection{Temperature dependency} \label{sec.ising_temperature_dependency}
\begin{figure}[t!]
\centering
\includegraphics[width=1.0\columnwidth]{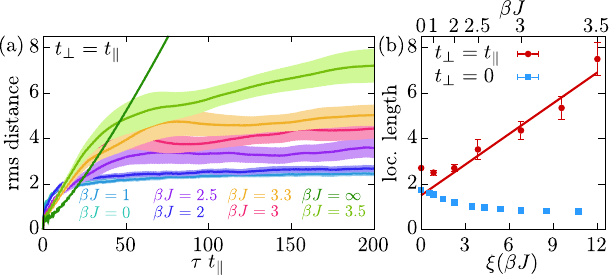}
\vspace{-0.5cm}
\caption{\textbf{Temperature dependence of localization.} (a) Dopant rms dynamics for indicated inverse temperatures, $t_\perp = t_\parallel$ and $J = 8t_\parallel$ showing persistent localization at any nonzero temperature. Moreover, a ballistic regime appears at intermediate timescales $\tau t_\parallel \sim 10$ for the lowest nonzero temperatures. This indicates the emergence of quasiparticles, here with lifetimes $\sim 20/t_\parallel$. Shaded areas show estimated standard errors on the mean from $200$ spin samples. For the TEBD, we use a truncation error of $10^{-8}$ and system sizes ranging from $N_x \times N_y = 45\times 2$ ($\beta J = 1$) to  $N_x \times N_y = 161\times 2$ ($\beta J = 3.5$). We note that the line for $\beta J = 0$ is almost entirely hidden by the $\beta J = 1, 2$ lines.   (b) Corresponding localization length for indicated $t_\perp$ vs the inverse temperature (top axis) as well as the underlying spin-spin correlation length $\xi(\beta J)$ (bottom axis), revealing a linear dependency (red line) for $t_\perp = t_\parallel$ and a saturating localization length (blue line) for $t_\perp=0$.}
\vspace{-0.25cm}
\label{fig:temperature_dependence_ising}
\end{figure}

We end our analysis of the Ising limit by investigating how susceptible the localization effect is to lowering the temperature away from $T=\infty$. To this end, we perform classical Monte Carlo sampling of the thermal spin background and again calculate the subsequent quantum dynamics using TEBD, as described by Eq.~\eqref{eq.monte_carlo_sampling_dynamics}. The resulting rms dynamics of the dopant is given in Fig.~\ref{fig:temperature_dependence_ising}(a) in the case of $t_\perp = t_\parallel$, which shows persistent localization for all the investigated nonzero temperatures. We also note that at intermediate timescales a ballistic regime appears with a propagation speed in good agreement with the zero-temperature limit. Long-lived quasiparticles (magnetic polarons), hereby, emerge in the low-temperature regime, which, however, eventually dampen out for all the investigated nonzero temperatures and gives way to localized behavior on longer timescales. For $t_\perp > 0$, the associated localization length [Fig.~\ref{fig:temperature_dependence_ising}(b)],
\begin{align}
l_{\rm loc} = \lim_{\tau\to\infty} d_{\rm rms}(\tau),
\end{align}
shows linear scaling with the underlying correlation length 
\begin{align}
\!\!\!\!\xi(\beta J) = \Bigg[& -\frac{\beta|J|}{4} + \ln \Bigg( \coth\frac{1}{2}\beta |J| \, \cosh \frac{1}{4}\beta |J|  \nn \\
&+ \sqrt{\left(\coth \frac{1}{2}\beta |J| \, \cosh \frac{1}{4}\beta |J|\right)^2 - 1}\Bigg)\Bigg]^{-1}\!\!\!\!
\label{eq.correlation_length}
\end{align}
of the spin-spin correlator: $\braket{\hat{S}^{z}_\bi  \hat{S}^{z}_{\bi+\bd}} \sim \exp(-|
\bd|/\xi)$ \cite{Nielsen2023_3}. This is in stark contrast to the behavior for $t_\perp = 0$, where the localization length \emph{decreases} with decreasing temperature [Fig.~\ref{fig:temperature_dependence_ising}(b)]. For $t_\perp = 0$, this decrease in localization length is a result of the fact that the dopant to a larger and larger extent experiences the strongly localizing linear potential in Fig.~\ref{fig:zero_vs_inf_temp_tperp}(b). \\

To understand the linear scaling of the localization length with correlation length for $t_\perp > 0$, note that the dopant spreads out more and more efficiently via the Trugman loops in Fig.~\ref{fig:zero_vs_inf_temp_tperp}(c). This naturally leads to a localization length on the order of $\xi(\beta J)$, because $\xi(\beta J)$ sets the length scale over which the N{\'e}el antiferromagnetic pattern is upheld. In turn, it also sets the length scale over which the dopant experiences the periodic potential in Fig.~\ref{fig:zero_vs_inf_temp_tperp}(c). The results thus strongly suggest that the dopant delocalizes asymptotically with $\xi(\beta J) \sim \exp(\beta J)$ as $T \to 0$.  We note that a completely analogous scaling of the localization length with $\xi(\beta J)$ was previously found for \emph{ferromagnetic} spin interactions in the limit of $t_\perp = 0$ \cite{Nielsen2023_3}, whose origin is similar in that the dopant becomes more and more free to move at decreasing temperatures. 

\begin{figure*}[ht!]
\centering
\includegraphics[width=1.0\textwidth]{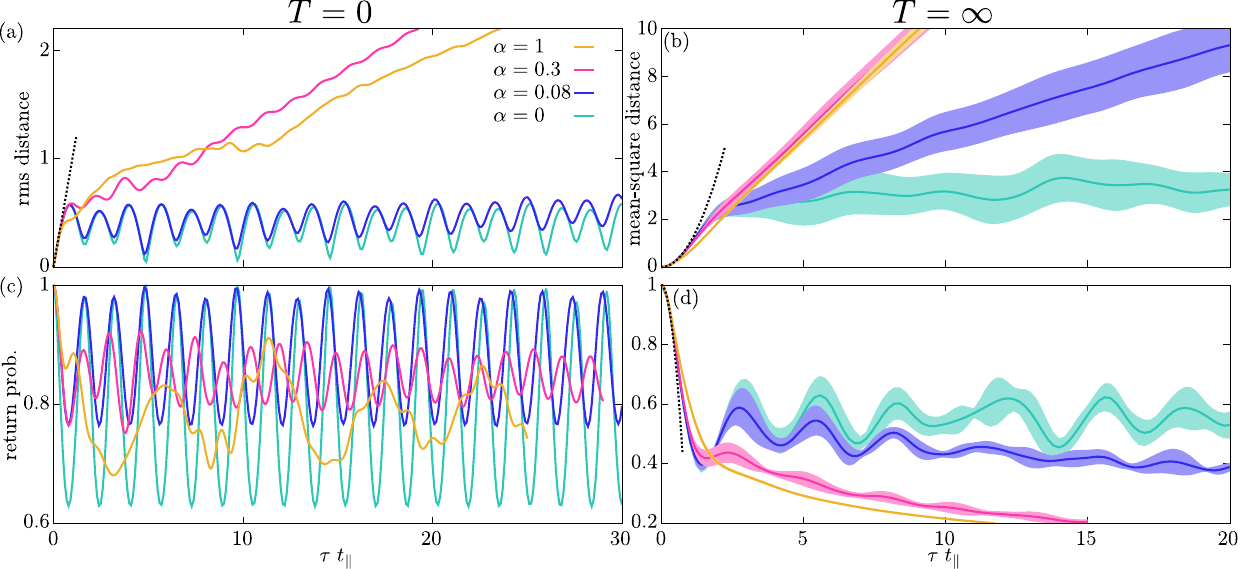}
\vspace{-0.5cm}
\caption{\textbf{Dopant dynamics vs spin exchange.} (a) Root-mean-square dynamics of the dopant at zero temperature for indicated values of the spin exchange, $\alpha$, and $J = 5t_\parallel$, showing universal ballistic behavior at short times (short-dashed line) and $\alpha$-dependent ballistic behavior at long times due to the formation of magnetic polaron quasiparticles. (b) Mean-square dynamics at infinite temperature for the same set of parameters. Here, we plot the mean-square distance (rather than the rms in (a)) to make the long-time linear diffusive behavior for any nonzero $\alpha$ apparent. This also means that the initial ballistic behavior is now a parabola. (c),(d) Corresponding probability for the dopant to return to its origin, $n_h({\bf i} = {\bf 0},\tau)$. Short-dashed lines show the initial ballistic behavior, where $d_{\rm rms}(\tau) = \tau\,t_\parallel$. Shaded areas show the estimated standard errors on the mean from 100 spin samples. For the TEBD, we use a truncation error of $10^{-7}$ and system sizes of $N_x\times N_y = 15\times 2, 21\times 2$ at low and high $\alpha$ at $T = 0$, whereas $N_x\times N_y = 11\times 2$ for $T = \infty$.}
\vspace{-0.25cm}
\label{fig:zero_vs_inf_temp_alpha}
\end{figure*}

\section{Spin exchange} \label{sec.spin_exchange}
We now turn to analyze the dopant motion once the spin exchange is turned on ($\alpha > 0$), both with and without rung hopping. To achieve long enough evolution times at nonzero temperatures, we will mostly use  intermediate system sizes of $N_x\times N_y = 11\times 2$. Throughout our analysis, we have checked that the probability to be on the right edge remains below $0.01$, such that finite-size effects may still safely be ignored. Moreover, we continue to fix the maximum truncation error rather than the bond dimension to achieve highly accurate and unbiased results.

Fig.~\ref{fig:zero_vs_inf_temp_alpha}, hereby, compares the dynamics at zero ($T = 0$) and infinite ($T = \infty$) spin temperatures in the absence of rung hopping ($t_\perp = 0$). At $T = 0$ and for $0.2\leq \alpha \leq 1$, we see a clear three-stage process in qualitative agreement with the 2D behavior found using approximate diagrammatic techniques \cite{Nielsen2022_2}. The three-stage process consists of: (1) the dopant moves with the universal initial speed $v = t_\parallel$ [short-dashed line in Fig.~\ref{fig:zero_vs_inf_temp_alpha}(a)], (2) an intermediate oscillatory regime, before (3) a final ballistic regime sets in. The latter ballistic regime is due to the formation of magnetic polaron quasiparticles \cite{White2015}. Moreover, similar to the nonzero rung hopping case in the Ising limit, this ballistic behavior is accompanied by a remaining localized core at least at the intermediate timescales investigated here, as seen in the return probability in Fig.~\ref{fig:zero_vs_inf_temp_alpha}(c). 

At infinite temperatures, however, there is a stark qualitative contrast to the nonzero rung hopping case in the Ising limit: after the initial universal regime with propagation speed $v = t_\parallel$, the dopant is now seen to delocalize on long timescales [Fig.~\ref{fig:zero_vs_inf_temp_alpha}(b)]. Note that we here plot the \emph{mean-square} distance, i.e. $[d_{\rm rms}(\tau)]^2$ from Eq.~\eqref{eq.root_mean_square_distance}. After just a few hopping times, we see clear linear behavior,and hence the dynamics is diffusive for any nonzero $\alpha$, with $[d_{\rm rms}(\tau)]^2 \propto \tau$. This is also reflected in the decay of the return probability [Fig.~\ref{fig:zero_vs_inf_temp_alpha}(d)]. Nevertheless, the overall dependency of the transport on the spin exchange at zero and infinite temperature look similar. 

\subsection{Comparison to spin dynamics} 
Before we analyze the phenomena above in more detail, we take a step back and compute the spin dynamics generated by Hamiltonian $H_J$ in the absence of dopants to enable a direct comparison between the behavior of the charge (the dopant) and the spins at infinite temperature. 

The total spin-$z$ conserving dynamics of only spins is naturally encoded in the two-time spin correlator $S(x,\tau) = \braket{\hat{S}^{z}(x,\tau)\hat{S}^{z}(x_0,0)}$, with the operator $\hat{S}^{z}(x,\tau) = \exp(i\Ham_J \tau) \hat{S}^{z}(x,0) \exp(-i\Ham_J \tau)$ in the Heisenberg picture and $\hat{S}^{z}(x,0) = 2\sum_{y={0,1}} \hat{S}^{z}((x,y),0)$ the sum of spin operators on rung $x$ over both legs. Here, the origin rung $x_0$ is placed in the middle of the ladder rather than at the left edge, as we are less limited by system size as compared to dopant simulations. The corresponding rms distance of the spin dynamics is then given by
\begin{align} \label{eq.spin_rms_dynamics}
d^s_{\rm rms}(\tau) = \Bigg[ \sum_x (x-x_0)^2 S^z(x,\tau) \Bigg]^{1/2},
\end{align}
so that $[d_{\rm rms}^s]^2$ is the width of a spreading wave packet of spin placed on the origin rung amidst a random spin background. To compute the spin correlator, we use the density matrix truncation (DMT) algorithm~\cite{White_2018}, a modified MPS time evolution method designed to efficiently capture local properties of time-evolved density matrices of bosonic systems. We use a recently introduced variant for Heisenberg evolving operators beyond one dimension, such as our two-leg ladder system, with a 1D MPS ansatz. This algorithm enables us to go to considerably larger system sizes ($501 \times 2$) and timescales ($J\tau \sim 1600$) for the spin dynamics than for hole dynamics, which we find to be crucial to get accurate results \cite{anand2025}, as we also comment further on below. The resulting rms dynamics is shown in Fig.~\ref{fig:spin_vs_charge}(a) for the indicated values of the spin anisotropy. We are, hereby, in a position to directly compare the observed {diffusive dynamics,
\begin{align} \label{eq.diffusive_behavior}
[d^h_{\rm rms}(\tau)]^2 &\to 2D_h (\tau - \tau^*_h), \nn \\
[d^s_{\rm rms}(\tau)]^2 &\to 2D_s (\tau - \tau^*_s),
\end{align}
for the hole (top) and spins (bottom). Note that since we start from a localized excitation, the initial dynamics is ballistic. This introduces time offsets $\tau^*_h, \tau^*_s$ [Fig.~\ref{fig:spin_vs_charge}(b)] in the diffusive behavior. While one might na{\"i}vely fit a straight line to the log-transformed data [$\ln(d^h_{\rm rms}(\tau))$ vs $\ln(\tau)$] to get the asymptotic dynamics, the large time offsets at weak spin exchange lead to artificial subdiffusive fits, in which $d^h_{\rm rms} \sim \tau^{\nu}$ with $\nu < 1/2$. A similar fitting problem for finite-time data was recently pointed out in understanding the universal coarsening dynamics of weakly interacting Bose gases~\cite{Gazo2025}. We give an example of this subtlety in Appendix \ref{app:subdiffusive_fits}.

We extract the diffusion coefficients $D_h, D_s$ in Eq.~\eqref{eq.diffusive_behavior} from our diffusive fits, as shown in Fig.~\ref{fig:spin_vs_charge}(c). We find that both $D_h$ and $D_s$ increase linearly for weak spin exchange ($\alpha \ll 1$). Additionally, at strong spin exchange ($\alpha > 1$), we find that the diffusion coefficient for the dopant begins to decrease again as $\alpha^{-1}$, while the spin coefficient crosses over to a faster linear behavior in this regime. The latter behavior is expected from the well-studied XX model, which we approach by taking $\alpha \gg 1$. For the XX ladder, previous studies have found $D_s \simeq 0.95 \alpha J$~\cite{Steinigeweg2014,Kloss2018,Rakovszky2022}, which is in good agreement with our results for $\alpha > 1$ [blue line in Fig.~\ref{fig:spin_vs_charge}(c)]. In this manner, there is \emph{qualitative} agreement between the behaviors of the spin and charge degrees of freedom. On the other hand, the dichotomy between the spins and the charge in their \emph{quantitative} behavior shows that the charge is not simply diffusing with the underlying spins. This is especially evident in the behavior for $\alpha > 1$, where the spins diffuse faster, while the dopant diffuses slower. We also note that while the diffusive behavior for the dopant sets in already after a few hopping times, $ t_\parallel \tau \sim 3$ (or $J \tau \sim 15$), on very short length scales, the spin dynamics needs around $J\tau \sim 100$ and considerably larger length scales. This a posteriori justifies our use of intermediate system sizes for the dopant dynamics, whereas we need much larger system sizes and timescales for the spin dynamics. 

{Finally, in Fig.~\ref{fig:spin_vs_charge}(d) we show the long-time ballistic propagation speed, $v_p$, at zero temperature obtained from linear fits to the rms dynamics in Fig.~\ref{fig:zero_vs_inf_temp_alpha}, arising due to the formation of magnetic polaron quasiparticles~\cite{White2015}. The dependency of this speed for $T=0$ on $\alpha$ qualitatively behaves the same as $D_h$ for $T=\infty$, with a maximum of $v_p$ around $\alpha = 0.4$. Physically, however, we definitely \emph{do not expect} the high-temperature regime to be described by Brownian motion of these magnetic polarons. Indeed, when the spins are completely disordered, the system does not even exhibit short-range antiferromagnetic order. Therefore, there should be no remnant of the low-temperature quasiparticles left. 

{Indeed, if the diffusion consisted of magnetic polarons undergoing Brownian motion, we should expect the Einstein-Smoluchowski expression \cite{Chandrasekhar1943,Islam2004} to be upheld, i.e. $D_h \propto v_p l$ with the mean free path $l$. Moreover, since the spins are completely disordered we should expect the mean free path to be on the order of the lattice spacing, $l \sim 1$. However, since $D_h \sim \alpha$ and $v_p \sim \alpha^2$ for low $\alpha$, while $D_h \sim \alpha^{-1}$ and $v_p \sim \alpha^{-4}$ for large $\alpha$, this simple relation does not seem possible. We note that while these power-law behaviors are only indicative, they do show that the propagation speed, $v_p$, at zero temperature behaves \emph{quantitatively} different from the infinite temperature diffusion coefficient, $D_h$. In this sense, high-temperature diffusion does not seem compatible with mere quasiparticle Brownian motion, but arises due to more complex many-body effects.

{The reason that $v_p(\alpha)$ and $D_h(\alpha)$ still behaves qualitatively the same can be understood from the fact that both very low and very high values of $\alpha$ leads to higher energy cost of producing spin excitations in the system than at intermediate values of $\alpha$. This suppresses transport -- both at low and high temperatures. 

\begin{figure}[t!]
    \centering
    \includegraphics[width=1.0\columnwidth]{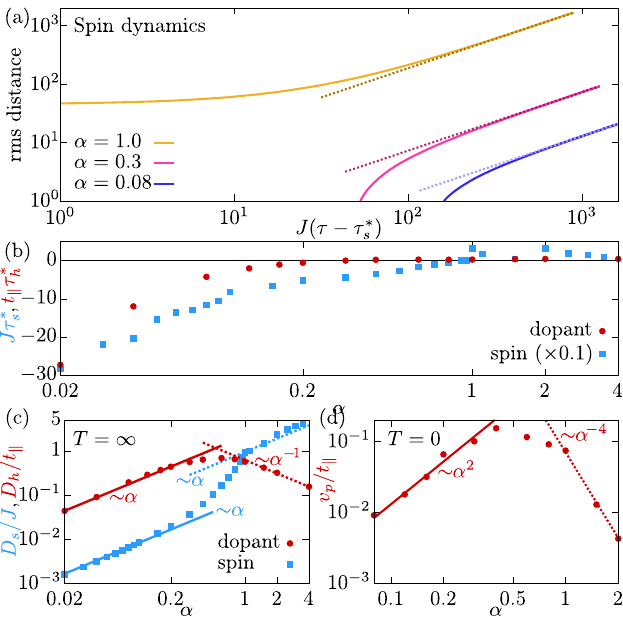}
    \vspace{-0.5cm}
    \caption{\textbf{Spin vs charge dynamics.} (a) Rms spin dynamics at infinite temperature for indicated values of the spin anisotropy, $\alpha$, in the absence of dopants along with diffusive fits (dashed lines). Note that time is now in units of the spin coupling, $J$, and that we include an $\alpha$ dependent time offset $\tau^*_s$ extracted from linear regression of $[d_{\rm rms}(\tau)]^2 \sim 2D_s (\tau - \tau^*_s)$ vs $\tau$. System sizes between $151 \times 2$ and $501 \times 2$ are used, and data end either at the maximal evolution time or when the boundary effects become non-negligible ($S(0,\tau) / S(x_0, \tau) > 10^{-3}$). (b) Comparison of the time offsets in the long-time diffusive behavior [Eq. \eqref{eq.diffusive_behavior}] for the spins and dopant [the former has been multiplied by $0.1$ to make the scales comparable]. (c) Diffusion coefficients, $D_s,D_h$, for the spins and dopant. For both, we find a linear onset $\sim \alpha$ [solid lines]. At strong spin exchange, the spins follow the expected linear behavior [dashed blue line] from the XX ladder with $D_s \simeq 0.95\alpha J$, whereas the diffusion coefficient for the dopant {decreases, seemingly following a $\sim \alpha^{-1}$ drop off} [dashed red line]. (d) Ballistic polaronic propagation speed $v_p$ vs $\alpha$ at zero temperature. This seems to follow a quadratic onset at weak spin exchange [solid red line] and a quartic drop-off at strong spin exchange [dashed red line]. We use $J = 5t_\parallel$  and $t_\perp = 0$ as in Fig.~\ref{fig:zero_vs_inf_temp_alpha}. }
    \vspace{-0.2cm}
    \label{fig:spin_vs_charge}
\end{figure}

\subsection{Self-similar scaling}

\begin{figure}[t!]
    \centering
    \includegraphics[width=1.0\columnwidth]{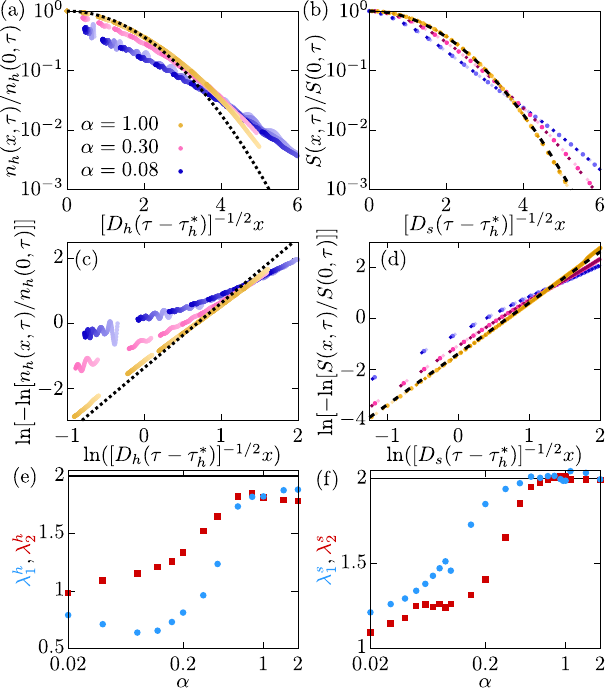}
    \vspace{-0.5cm}
    \caption{\textbf{Self-similar scaling behavior.} Scaled hole distribution (a) and scaled spin-spin correlator (b) as a {function of $u_i = [D_i(\tau - \tau^*_i)]^{-1/2} x$} for indicated spin anisotropies $\alpha$ reveals self-similar scaling behavior. We compare these to the na{\"i}vely expected Gaussian distribution $\sim \exp[-(u_i/2)^2]$ from regular diffusion shown in short-dashed black lines. The color gradient on the data points denote the evolution time from shortest used (bright colors) to longest used (dark colors). The used intervals are: (a) $t_\parallel\tau  \in [6,20], [6,11], [6,11]$ and (b) $J\tau  \in [1400,1600], [1000,1200], [720,920]$ for $\alpha = 0.08,0.3,1.0$. (c),(d) The double-logarithm of the scaled distribution vs $\ln([D_i(\tau - \tau^*_i)]^{-1/2} x)$ shows differing \emph{linear} behaviors on short and long scales with slopes $\lambda_1,\lambda_2$ respectively. This describes generalized exponential distributions, $n_h\sim \exp[-(\tau^{-1/2} x)^{\lambda^h}]$, $S^z\sim \exp[-(\tau^{-1/2} x)^{\lambda^s}]$. Again short-dashed lines shown the expected Gaussian behavior. The fitted general exponents $\lambda_1,\lambda_2$ at short and long scales for the hole distribution (e) and spin-spin correlator (f) are shown in blue and red, respectively, against the spin anisotropy, $\alpha$. For low $\alpha$, these \emph{strongly} deviate from diffusive behavior (black lines: $\lambda = 2$).}
    \vspace{-0.25cm}
    \label{fig:self_similar}
\end{figure}

To elucidate the diffusion process further, we investigate whether simple hydrodynamic diffusion equations can explain the dynamics. That is, is it possible to describe the density and spin distribution dynamics in terms of emergent hydrodynamic equations, $\partial_\tau n(x,\tau) = D_h\,\partial_x^2 n(x,\tau)$ and $\partial_\tau S(x,\tau) = D_s\,\partial_x^2 S(x,\tau)$ for the rung hole distribution $n_h(x,\tau) = n_h((x,0),\tau) + n_h((x,1),\tau)$ and the rung spin correlator $S(x,\tau)$, respectively? If this were the case, the dynamics should show self-similar scaling {(including the time offset $\tau^*_h, \tau^*_s$)}, 
\begin{align} \label{eq.scaling_behavior}
\!\! n_h(x,\tau) &= \frac{1}{[D_h (\tau - \tau^*_h)]^{1/2}} F_h\left(\frac{x}{[D_h (\tau - \tau^*_h)]^{1/2}}\right), \!\!\nn \\
\!\! S(x,\tau) &= \frac{1}{[D_h (\tau - \tau^*_s)]^{1/2}} F_s\left(\frac{x}{[D_h (\tau - \tau^*_s)]^{1/2}}\right),\!\!
\end{align}
with a \emph{Gaussian} scaling function $F_h(u) = F_s(u) = \exp(-u^2/4)$ \cite{Crank_1975}. We check this behavior in Figs.~\ref{fig:self_similar}(a) and \ref{fig:self_similar}(b) and indeed find robust self-similar scaling. Notably however, the scaling behavior \emph{strongly} deviates from Gaussianity, especially for weak spin exchange. To make this more apparent, we replot the data on a log double-log plot in Figs.~\ref{fig:self_similar}(c) and \ref{fig:self_similar}(d). This reveals distinct linear behaviors at short and long scales, with slope $\lambda_1$ and $\lambda_2$ respectively. It follows that the scaling functions take on the limiting short- and long-range behaviors
\begin{align}
F_i(u) \propto
\begin{cases}
\exp\Big[-\Big(\frac{u}{\sqrt{2}\sigma_1^{i}}\Big)^{\lambda_1^{i}}\Big], & \; u \lesssim 1, \\
\exp\Big[-\Big(\frac{u}{\sqrt{2}\sigma_2^{i}}\Big)^{\lambda_2^i}\Big], & \; u \gtrsim 1,
\end{cases}
\end{align}
{where $u = x / [D_i(\tau - \tau^*_i)]^{1/2}$ is the rescaled coordinate}, and $i = h, s$ indexes whether it is for the hole or spin degree of freedom. These generalized exponential functions, hereby, come with emergent exponents $\lambda_1^i,\lambda_2^i$ on short and long scales, which we plot over a wide range of spin anisotropies in Figs.~\ref{fig:self_similar}(e) and \ref{fig:self_similar}(f). For large enough $\alpha$, these exponents are similar to one another and for the spin degree of freedom it falls onto the expected Gaussian behavior $\lambda = 2$. For the dopant, however, even in this limit of fast spin exchange, the scaling function still deviates from Gaussianity with $\lambda^h_1 = \lambda^h_2 \simeq 1.8$. More dramatically, at lower $\alpha$ they both start to deviate substantially from Gaussianity and each other. Interestingly, the charge and spin degree of freedom show opposite tendencies here. Whereas the hole distribution develops a cusp on short scales, with $\lambda_1^h<\lambda_2^h$, the spin distribution is soft at short scales and sharper at longer scales, $\lambda_1^s>\lambda_2^s$. Once again, this emphasizes the qualitative agreement between the spin and charge degrees of freedom, albeit with dichotomous quantitative behavior. 

The appearance of such self-similar scaling is evidence that the system has settled into pre-thermal behavior \cite{Calabrese2005,Chandran2013,Maraga2015,Feldmeier2020,Joaquin2022} associated with non-thermal fixed points \cite{Berges2008,Bhattacharyya2020,Mikheev2023}. In this context, the time offsets $\tau^*_h,\tau^*_s$ contain the only remaining dependency on the initial state, whereby the universal diffusive dynamics depends on the ``universal clock time'' $\tau - \tau^*$ \cite{Gazo2025}.

The strong deviation away from the Gaussian behavior at low $\alpha$, which is otherwise inherent to diffusion, indicates that the dynamics actually cannot be described by simple diffusion equations, even though the second moment, $[d_{\rm rms}(\tau)]^2$ increases linearly in time. Such \emph{non-Gaussian diffusion} has been observed in a wide range of soft matter systems~\cite{Weeks2000,Hapca2009,Wang2009,Toyota2011,Parry2014,Munder2016}, and appears classically because the individual steps of the diffusing particles are \emph{not} independent and identically distributed. One possible way of modeling this behavior is by so-called \emph{superstatistics}~\cite{Chechkin2017,Metzler2020}, in which the diffusion coefficient itself is drawn from a probability distribution, $p(D)$, e.g. due to ensembles with varying diffusion coefficients. The ensemble averaged distribution in such a framework is, thus, described by $G_s(x,\tau) = \int_0^\infty dD \, p(D) g(x, \tau | D)$, in which $g(x, \tau | D)$ is the regular Gaussian distribution with diffusion coefficient $D$. This model has three key properties: (1) the second moment remains linear, $\braket{x^2} = 2\braket{D} \tau$, where $\braket{D} = \int_0^\infty dD \,p(D) D$ is the ensemble-averaged diffusion coefficient, (2) $G_s(x,\tau)$ retains self-similar scaling behavior, and (3) the associated scaling function can deviate strongly from Gaussianity} \footnote{In Refs. \cite{Chechkin2017,Metzler2020}, they give an experimentally relevant example of the exponential distribution $p(D)$, which gives rise to an exponential scaling function $F(u) \sim \exp[-|u|]$.}. In our system, the ensemble of diffusion coefficients could potentially arise from the different microscopic realizations of the thermal spin ensemble. This is supported in the hole transport numerics by the fact that the strong deviations away from Gaussianity at low $\alpha$ is accompanied by a larger spread in $d_{\rm rms}(\tau)$ over the $100$ spin samples -- i.e. the standard error on $d_{\rm rms}(\tau)$ is larger at low $\alpha$ as seen directly in Fig.~\ref{fig:zero_vs_inf_temp_alpha}(b).

A distinct possibility is, however, that the Gaussian behavior emerges at sufficiently long times, as e.g. described in classical \emph{diffusing diffusivity} models \cite{Chubynsky2014,Chechkin2017,Metzler2020}, such that the non-Gaussian behavior is a transient behavior only. Naturally, we cannot determine whether this is the case. However, in our finite-time data we do not see any hints of the distribution skewing towards a Gaussian shape as a function of time.

\subsection{Rung hopping} \label{sec.spin_exchange_rung_hopping}

\begin{figure}[t!]
    \centering
    \includegraphics[width=1.0\columnwidth]{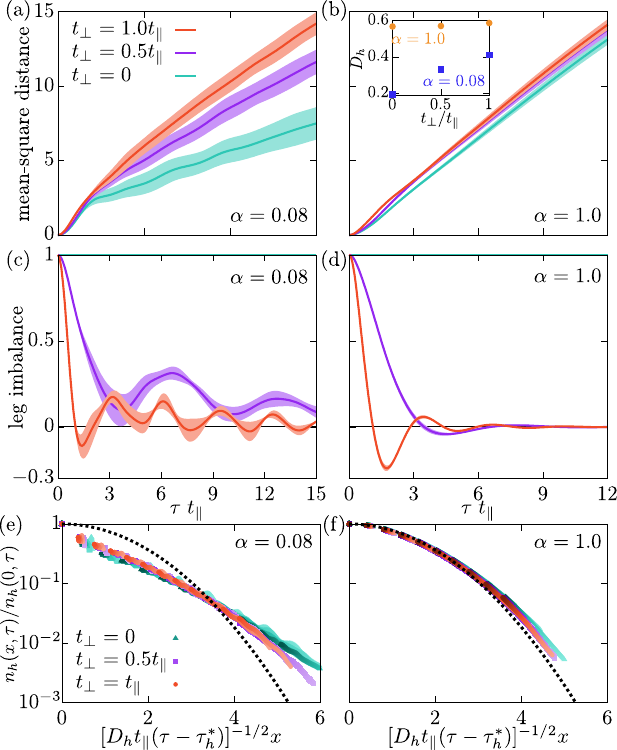}
    \vspace{-0.5cm}
    \caption{\textbf{Rung hopping dependency.} [(a) and (b)] Dopant mean-square dynamics at indicated spin anisotropies, $\alpha$, and transverse hopping, $t_\perp$. While the associated diffusion coefficients [inset] increases with $t_\perp$ for $\alpha = 0.08$, it is basically constant for $\alpha = 1$. [(c) and (d)] The leg imbalance [Eq.~\eqref{eq.leg_imbalance}] shows that the legs of the ladder thermalize on a timescale depending on $t_\perp$ and $\alpha$. Shaded areas show the estimated standard error from 100 spin samples. [(e) and (f)] Hole distributions for indicated hopping and spin-exchange parameters, showing a very weak dependency on $t_\perp$. The self-similar scaling, hereby, retains their deviation from a purely Gaussian distribution $\sim \exp[-(u/2)^2]$ (short-dashed black lines). For the simulations, we use a truncation error of $10^{-7}$ and $N_x\times N_y = 11\times 2$. }
    \vspace{-0.25cm}
    \label{fig:spin_exchange_rung_hopping}
\end{figure}

To complete our understanding of the dynamical phase diagram in Fig.~\ref{fig:HSF}(d), we now investigate whether the discovered non-Gaussian diffusion is a result of the proximity to the \emph{fragmented phase} at $\alpha = t_\perp = 0$. To investigate this, we track the hole motion at nonzero rung hoppings, $t_\perp / t_\parallel = 0.5, 1$, at spin anisotropies $\alpha = 0.08$ and $1$. The analysis shown in Figs.~\ref{fig:spin_exchange_rung_hopping}(a) and \ref{fig:spin_exchange_rung_hopping}(b) reveals that the diffusive behavior at $\alpha = 1$ is robust to changes in the transverse hopping $t_\perp$ and that the long-time behavior of $t_\perp = 0.5t_\parallel$ and $t_\perp = t_\parallel$ collapses on longer timescales. For $\alpha = 0.08$ on the other hand, the diffusion coefficient increases by a factor of $2$ between $t_\perp = 0$ and $t_\perp = t_\parallel$ (see inset in Fig.~\ref{fig:spin_exchange_rung_hopping}(b)). Moreover, Figs.~\ref{fig:spin_exchange_rung_hopping}(e) and \ref{fig:spin_exchange_rung_hopping}(f) shows that the self-similar scaling function only depends very weakly on $t_\perp$. In particular, the distributions for $\alpha = 0.08$ retain their strong deviation from Gaussianity, whereas the distributions at $\alpha = 1$ retain their small deviation from Gaussianity well-described by the exponential distribution $\exp[-(x/\tau^{1/2})^\lambda]$ with $\lambda \simeq 1.8$. In this manner, the non-Gaussian diffusive behavior is \emph{not} a special property happening only close to the fragmented phase at $\alpha = t_\perp = 0$. \\

Finally, the presence of a nonzero $t_\perp$ gives us a natural way of probing \emph{local} thermalization of the system. In Figs.~\ref{fig:spin_exchange_rung_hopping}(c) and \ref{fig:spin_exchange_rung_hopping}(d), we show the leg imbalance
\begin{align}
    \calI(\tau)= \sum_{x} \left[ n_h((x,0),\tau) -  n_h((x,1),\tau) \right], 
    \label{eq.leg_imbalance}
\end{align}
measuring the difference in hole density in the two legs of the ladder. When $t_\perp = 0$, the hole cannot hop to the other leg, whereby $I_l = 1$. For $t_\perp > 0$, however, we see that the leg imbalance diminishes over time with heavily damped oscillations around $I_l = 0$, which indicates that the dopant eventually ``forgets" that it was initially localized on the $y = 0$ leg, showing the anticipated thermalization. This process is, however, significantly slower for small spin anisotropies. We note that in the extreme Ising limit of $\alpha = 0$, $I_l$ remains nonzero on all timescales, evidencing the localization-induced lack of thermalization described in Sec. \ref{sec.ising_limit}.

\subsection{Temperature dependency}  \label{sec.spin_exchange_temperature_dependency}
Let us now address how the system crosses over from its diffusive behavior at infinite temperature to ballistic behavior at zero temperature. To this end, and as explained in Sec.~\ref{sec.methods}, we use METTS to sample the thermal background of the spins. The exact methodology is detailed in Appendix \ref{app:METTS_details}, where we also show that the achieved ensembles show no discernible autocorrelations. The resulting mean-square dynamics are shown in Fig.~\ref{fig:temperature_dependence_alpha} for indicated values of the spin anisotropy and $t_\perp = 0$. We note that at a fixed time, the mean-square distance smoothly crosses over from its infinite to zero temperature value. Moreover, at high temperatures $\beta J = J / k_B T \leq 2$, we find that the diffusion coefficient $D_h$ at fixed $J/t_\parallel$ is very well described by an Arrhenius relation
\begin{equation}
    D_h(\alpha, \beta J) = D_h(\alpha,0) \exp\left[-\varepsilon(\alpha)\beta J \right],
    \label{eq.arrhenius_relation}
\end{equation}
in which $D_h(\alpha,0)$ is the infinite temperature diffusion coefficient. In this manner, $D_h$ decreases \emph{exponentially} with decreasing temperatures, as evident from Fig.~\ref{fig:temperature_dependence_alpha}(c). Arrhenius originally suggested this relation for molecular reaction rates at high temperatures \cite{Arrhenius1889}. It also gives an accurate description of impurity diffusivity in the bulk \cite{Katz1971,Peterson1978,Ishikawa1985} and on the surface of metals \cite{Bonzel1970} at high temperatures and may even be derived from rather general classical statistical-mechanical considerations \cite{Fujita1988}. It is, however, to the best of our knowledge, the first time that such a relation emerges from a first-principle quantum-mechanical description. The Arrhenius form also suggests that we can interpret $\varepsilon$ as an \emph{activation energy} of the diffusive process. We note that the activation energy is very similar for $\alpha = 0.08$ and $\alpha = 1$. Qualitatively, we would} expect the activation energy to be larger at larger $\alpha$, because the \emph{change in magnetic energy} in a single hopping event fluctuates more at larger $\alpha$. Indeed, for infinite temperatures this can be evaluated to yield fluctuations with a standard deviation of ${\rm std}(\Delta E_J) = \sqrt{1 + 2\alpha^2}J/2$, which indeed increases with $\alpha$ (see Appendix \ref{app:magnetic_energy_fluctuations} for details). This simple expression, however, significantly \emph{overestimates} the ratio of $\varepsilon(\alpha = 1) / \varepsilon(\alpha = 0.08) \simeq 1.13$ fitted from the data in Fig.~\ref{fig:temperature_dependence_alpha} by a factor of $\sim 1.5$. This discrepancy is presumably due to the fact that one should rather describe the process not at short distance scales but in terms of minute long-wavelength corrections to the thermal background, akin to what is done in generalized hydrodynamics \cite{Doyon2025}.   

\begin{figure}[t!]
    \centering
    \includegraphics[width=1.0\columnwidth]{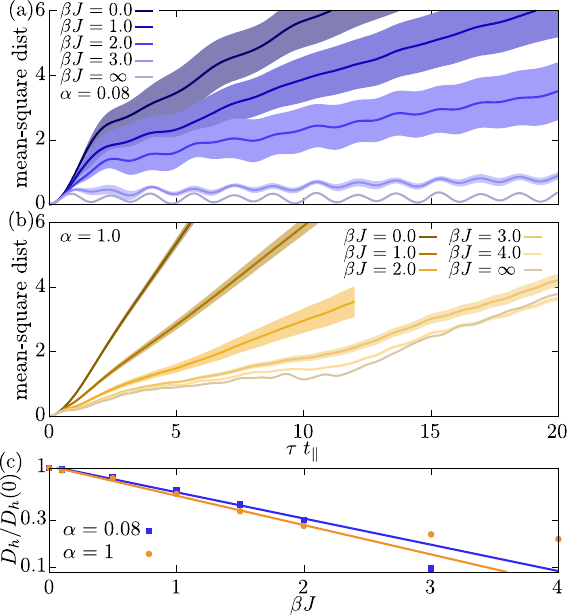}
    \vspace{-0.5cm}
    \caption{\textbf{Temperature dependency of diffusion.} Mean-square dynamics for indicated inverse temperatures, $\beta J$, and spin anisotropies $\alpha = 0.08$ (a), $\alpha = 1$ (b), at $J = 5 t_\parallel$ and $t_\perp = 0$. Shaded areas show the estimated standard error from 100 METTS samples. (c) Extracted diffusion coefficients as a function of inverse temperature on a linear-log scale, where colored lines show fits to the Arrhenius form in Eq.~\eqref{eq.arrhenius_relation}. For the TEBD, we use a truncation error of $10^{-7}$ and $N_x\times N_y = 13\times 2$.}
    \vspace{-0.5cm}
    \label{fig:temperature_dependence_alpha}
\end{figure}

The mere appearance of an Arrhenius relation indicates that we can think of the diffusion process semi-classically in the following way. When the dopant moves, it must cross an energy barrier given by the magnetic fluctuations. At infinite temperature, the dopant completely disregards this barrier and moves with a diffusion coefficient $D_h(\alpha,\beta J = 0)$. As the temperature is lowered, however, it becomes increasingly difficult for the dopant to cross the barrier leading to the exponential reduction in Eq.~\eqref{eq.arrhenius_relation}. We believe that this should hold whenever the spin couplings dominate the hopping process, $J / t \gg 1$. At intermediate and low values of $J / t$, this simple picture could break down. However, this regime is technically challenging to access, because the dopant in this case has a prolonged initial ballistic regime, and so we would need to do computations for increasingly large spin lattices. 

We note that at sufficiently low temperatures, we expect deviations from this behavior. Indeed, this regime should eventually be described by long-lived quasiparticles with transient ballistic behavior. We see the first indications of such behavior in the data at inverse temperature $\beta J = 4$ for $\alpha = 1$, in which it follows the zero temperature behavior until $\tau = 15 / t_\parallel$. Moreover, since the dopant propagates ballistically at zero temperature at long times, it must eventually cross all the mean-square distance curves for nonzero temperatures ($\beta J < \infty$). The first signs of this are indeed seen in Fig.~\ref{fig:temperature_dependence_alpha}(b), where the zero-temperature curve crosses the one for $\beta J = 4$ around $\tau = 17 / t_\parallel$.

\subsection{Spin coupling dependency} \label{sec.spin_exchange_coupling_dependency}
\begin{figure}[t]
    \centering
    \includegraphics[width=1.0\columnwidth]{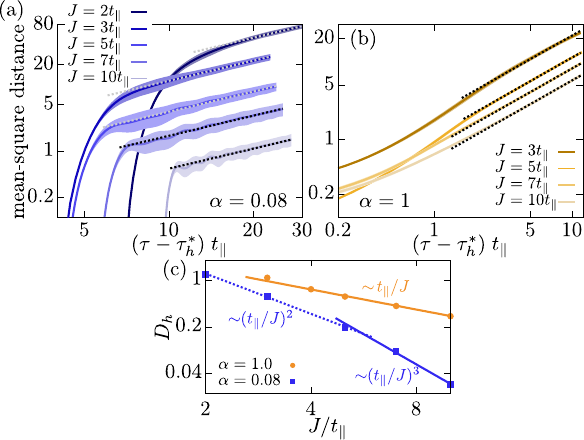}
    \vspace{-0.5cm}
    \caption{\textbf{Spin coupling dependency of diffusion.} Dopant mean-square dynamics for indicated values of the spin coupling, $J$, for very weak (a) and strong (b) spin exchange, both for $t_\perp = 0$ on a log-log plot. These are shifted by the time offset $\tau_h^*$ obtained from the linear fits $[d_{\rm rms}^h(\tau)]^2 = 2D_h (\tau - \tau_h^*)$ to reveal that all dynamics eventually follow diffusive behavior (dashed lines). Shaded areas show the estimated standard error from 100 samples. (c) Diffusion coefficient as a function of $J/t_\parallel$. The dashed and solid lines show several indicative power-law behaviors. For the TEBD, we use a truncation error of $10^{-7}$. For $\alpha = 1$, $N_x\times N_y = 13\times 2$, whereas for $\alpha = 0.08$ it ranges from $N_x\times N_y = 13\times 2$ at $J = 10t_\parallel$ to $N_x\times N_y = 31\times 2$ at $J=2t_\parallel$ to capture the faster increase in the rms dynamics as $J/t_\parallel$ is lowered.}
    \vspace{-0.5cm}
    \label{fig:alpha_J_dependency}
\end{figure} 

In this section, we investigate quantitatively the dependency of the infinite temperature dopant dynamics on the ratio between the spin coupling and the hopping amplitude, $J/t_\parallel$. The resulting dopant mean-square dynamics are shown in Figs.~\ref{fig:alpha_J_dependency}(a) and \ref{fig:alpha_J_dependency}(b), showing \emph{faster} delocalization at \emph{smaller} spin couplings. The associated diffusion coefficient, hereby, \emph{decreases} for increasing spin coupling, once again emphasizing that the charge does not simply follow the diffusing spins. In contrast, for fixed $\alpha$, the spin diffusivity -- in the absence of holes -- \emph{increases} linearly in $J$, as this is the only energy scale in this case. The reason for the decreasing dopant diffusion coefficient is that higher spin couplings leads to larger energy costs for producing spin excitations. As such excitations are inevitable created as the dopant moves, higher $J$ suppresses transport.

More quantitatively, in the investigated regime the diffusion coefficients at $\alpha = 1$ follows a $D_h\sim t_\parallel / J$ behavior for $J > 3t_\parallel$. On the other hand, the diffusion coefficient at $\alpha = 0.08$ shows a much stronger suppression for larger $J/t_\parallel$ and seems more compatible with $D_h\sim (t_\parallel / J)^3$ at large $J/t_\parallel$. Finally, while we do not show this explicitly in Fig.~\ref{fig:alpha_J_dependency}, we note that the non-Gaussian self-similar scaling, investigated in Fig.~\ref{fig:self_similar} at $J/t_\parallel = 5$, only depends very weakly on $J/t_\parallel$, indicating the robustness of this behavior.

\section{Towards two-dimensional lattices} \label{sec.towards_2D}
In this section, we discuss how our results may generalize to a fully two-dimensional system. We start in the Ising limit, $\alpha = 0$, with isotropic hopping amplitudes $t_\perp = t_\parallel \equiv t$, for which we can make precise qualitative arguments. In this case, we hypothesize that the dopant is still localized at high temperatures, $\beta J \to 0$, and that this remains true even across the Ising phase transition at $\beta_c J = 2{\rm ln}[1 + 2^{-1/2}] \simeq 1.76$ \cite{Onsager1944}. To see why, we compute the magnetic energy landscape seen by the dopant as it travels through the Ising magnet, i.e. $E_J(\bsigma,l) = \bra{\psi_\bsigma(l)}\Ham_J\ket{\psi_\bsigma(l)}-\bra{\psi_\bsigma(0)}\Ham_J\ket{\psi_\bsigma(0)}$ in complete analogy with Fig.~\ref{fig:zero_vs_inf_temp_tperp}(h). In 2D, it may not be enough that this landscape is disordered to ensure dopant localization, since 2D is the lower critical dimension for Anderson localization \cite{Tarquini2017}. This is in contrast to Anderson localization in one dimension \cite{Anderson1958,Fouque1999}, as well as our present findings for the two-leg ladder in the Ising limit (see Fig.~\ref{fig:zero_vs_inf_temp_tperp}). However, if this landscape inevitably fluctuates more and more as a function of traveled distance of the dopant, this is a very strong argument for localization. In fact, if the magnetic energy at a traveled distance of $l$ has a standard deviation ${\rm std}[E_J(\bsigma,l)] = J \times (l/l_{\rm fl})^a$ for some power $a > 0$, the dopant should \emph{only} be able to travel out to distances where the fluctuations surpass its initial kinetic energy. This distance is set by $t \sim {\rm std}[E_J(\bsigma, l)]$, leading to a localization length on the order of $l_{\rm loc} \sim l_{\rm fl} (t/J)^{1/a}$ \cite{Nielsen2024_1}. 

To be more explicit, in the Ising limit where we can analyze classical spin configurations with one spin removed, we use classical Monte Carlo sampling of the spin background \cite{Hastings1970, Wolff1989, Nielsen2024_1} to produce $2000$ samples of size $N_x \times N_y = 201\times 201$ for a range of inverse temperatures and initialize the hole (the dopant) at the central site of the lattice. At low temperatures, the eventual delocalization of the dopant is enabled by Trugman loops \cite{Trugman1990} as discussed in Sec. \ref{sec.ising_limit} and shown in Fig.~\ref{fig:zero_vs_inf_temp_tperp}(a). In a single Trugman loop, the dopant travels along one of the four diagonals $(x,y) \to (x \pm 1, y \pm 1)$. Thus we can check how small the magnetic energy changes experienced by the dopant can possibly be by the following iterative process. At each site, we compute the \emph{change} in energy following one of the four Trugman loops. We then pick the site with the lowest energy change as the next starting point, update the spin background accordingly and repeat the procedure. To ensure that the dopant does not get stuck, we do \emph{not} allow it to retrace its own path. This will generate a path with the smallest possible change in magnetic energy from a local update rule and leads, in particular instances, to the paths shown in Figs. \ref{fig:towards_2D}(a) and \ref{fig:towards_2D}(b). We note that in Fig.~\ref{fig:towards_2D}(b), the system is below the critical temperature, $\beta > \beta_c$ and has spontaneously chosen the configuration with a majority of antiferromagnetic domains with staggered magnetization $M(x,y) = (-1)^{x+y} S^{z}(x,y) = -1/2$ (blue regions).

\begin{figure}[t!]
    \centering
    \includegraphics[width=1.0\columnwidth]{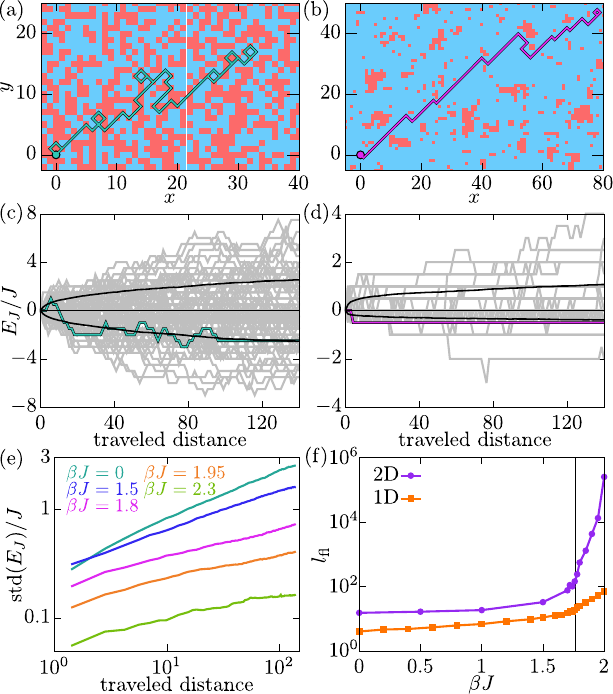}
    \vspace{-0.5cm}
    \caption{\textbf{2D motion in an Ising magnet.} Low-energy trajectories (colored lines) of a dopant in the paramagnetic phase at $\beta J = 0$ (a), and in the antiferromagnetic phase at $\beta J = 1.8$ (b) shown for a specific Monte Carlo sample of the spin background. The staggered spin pattern of this sample, $M(x,y) = (-1)^{x+y} S^{(z)}(x,y)$, is shown in the background in red [$M = +1/2$] and blue [$M = -1/2$]. [(c) and (d)] Magnetic energy vs traveled distance by the dopant is shown for the same inverse temperatures [(c): $\beta J = 0$, (d): $\beta J = 1.8$]  for $100$ Monte Carlo samples of the background [gray lines] and the particular ones used in (a) and (b) (colored lines). Black lines show the mean of $E_J$ plus/minus one standard deviation, $\mu(E_J)\pm {\rm std}(E_J)$, computed from $2000$ spin samples. (e) Standard deviation ${\rm std}(E_J)$ vs traveled distance of the dopant for indicated inverse temperatures. (f) Corresponding fluctuation length scale from power-law fit, ${\rm std}[E_J(l)]/J = (l/l_{\rm fl})^a$, for 2D (purple dots) and 1D (orange squares) motion of the dopant.}
    \vspace{-0.25cm}
    \label{fig:towards_2D}
\end{figure}

In this manner, the dopant travels more easily in the antiferromagnetic ($\beta > \beta_c$) phase than in the paramagnetic ($\beta < \beta_c$) phase. Indeed, in Figs.~\ref{fig:towards_2D}(c) and \ref{fig:towards_2D}(d), we show the corresponding change in magnetic energy, $E_J(\bsigma,l)$, as a function of traveled distance $l$ for $100$ of the $2000$ samples, $\bsigma$. We see that at infinite temperatures [Fig.~\ref{fig:towards_2D}(c)], this is symmetric around $E_J = 0$, whereas it becomes asymmetric and has much smaller fluctuations in the antiferromagnetic phase [Fig.~\ref{fig:towards_2D}(d)]. Moreover, we numerically find that the standard deviation of the magnetic energy, 
\begin{equation}
\!\!{\rm std}[E_J(l)]^2 = \frac{1}{|S|}\sum_{s\in S} E_J^2(s,l) - \Big(\frac{1}{|S|}\sum_{s\in S} E_J(s,l)\Big)^2,\!\!
\end{equation}
calculated from the Monte Carlo generated set $S$ of $|S| = 2000$ spin samples, retains a power-law behavior, ${\rm std}[E_J(l)] = J \times (l/l_{\rm fl})^a$, across the phase transition [Fig.~\ref{fig:towards_2D}(e)]. This is similar to the $l^{1/2}$ behavior described in Fig.~\ref{fig:zero_vs_inf_temp_tperp}(g) for $t_\perp = 0$, and simultaneously shows that the bounded behavior of ${\rm std}[E_J(l)]$ for $t_\perp > 0$ [Fig.~\ref{fig:zero_vs_inf_temp_tperp}(g)] is particular to the two-leg ladder. This behavior strongly suggests that the dopant is localized at infinite temperatures and that this \emph{remains true} across the phase transition, though with a larger localization length. In this manner, the associated fluctuation length scale $l_{\rm fl}$, with which the localization length should scale, is finite across the phase transition [Fig.~\ref{fig:towards_2D}(f)] and only shows a mild increase just around the critical temperature. At slightly lower temperatures (larger $\beta$), it does, however, increase much more rapidly  than in the case of 1D motion ($t_\perp = 0$) of a dopant in a 2D Ising magnet \cite{Nielsen2024_1}. Indeed, the increase in ${\rm std}[E_J(l)]$ at the lowest investigated temperatures of $\beta J = 2.3$ vs traveled distance $l$ becomes very slow and perhaps even levels off at large distances. Taken together, this indicates that the localization length remains finite across the phase transition, but becomes parametrically large as the temperature is decreased. Finally, since the fluctuation length scale $l_{\rm fl}$ seemingly increases exponentially below the phase transition, we can expect quasiparticles with correspondingly long lifetimes on the order of the time it takes them to travel to the localization length, $\sim l_{\rm fl} / v_p$, where $v_p$ is a typical polaronic propagation speed. 

A distinct possibility, however, is that below the phase transition, the dopant behaves as a charge scattering on static point like impurities. This scenario naturally arises in impure metallic conductors and in 2D gives rise to weak localization \cite{Dolan1979, Altshuler1980_1, Altshuler1980_2, BruusFlensberg}, in which the resistivity increases logarithmically at low temperatures. Indeed, one might think of the small $M = +1/2$ domains in Fig.~\ref{fig:towards_2D} as such static point-like impurities that the dopant is scattering off. Following this logic, the impurity density decreases with decreasing temperatures as the system becomes more antiferromagnetically ordered. If this is the correct physical picture, the dopant should go from being localized above the critical temperature to diffusing below the critical temperature. Moreover, as temperature decreases to zero long-lived quasiparticles should also emerge in this scenario, whereby the diffusion coefficient diverges in this limit, and dopants may finally move ballistically. 

Finally, in the presence of spin-exchange processes it is unclear whether the nature of the high-temperature behavior will change qualitatively as we go to two dimensions. Indeed, as a hint towards the full 2D behavior, we saw in Sec.~\ref{sec.spin_exchange_rung_hopping} that the inclusion of transversal hopping, $t_\perp > 0$, had no \emph{qualitative} impact on the dopant motion, as non-Gaussian behavior is still observed. Moreover, it seems that exactly the same processes will be available both in the ladder and in the full 2D case. However, as more legs are added to the ladder, it is possible that the strong deviation from Gaussian diffusion may eventually fade towards more conventional Gaussian transport. We give evidence of this phenomenology in Fig.~\ref{fig:multilegged_spin_ladder}, in which we investigate the spin dynamics in $151\times N_y$ cylindrical lattices for $N_y = 3,4$\footnote{$N_y=2$ is a strip with open boundary conditions in the short direction, while $N_y>2$ are cylinders with periodic boundary conditions in the short direction.}. In particular, we calculate the spin rung distribution $S(x,\tau) = \braket{\hat{S}^z(x,\tau)\hat{S}^z(x_0,0)}$, where we in analogy to our analysis in Sec.~\ref{sec.spin_exchange} define the rung operator $\hat{S}^z(x,0) = \sum_{y = 0}^{N_y}\hat{S}^z((x,y),0)$. From this, we calculate the rms distance dynamics according to Eq.~\eqref{eq.spin_rms_dynamics} for a weak spin exchange of $\alpha = 0.08$ for which we found deviations from Gaussianity for $N_y=2$. We find that diffusion persists in all three investigated cases [Fig.~\ref{fig:multilegged_spin_ladder}(a)] and that each retains a non-Gaussian self-similar scaling [Fig.~\ref{fig:multilegged_spin_ladder}(b)]. For the ``most'' two-dimensional case of $N_y = 4$, we additionally find that the scaling function is well-described by a sum of just two Gaussian distributions, 
\begin{equation}
F_s(u) = \sum_{j=1,2} C_j \exp\left[-\frac{u^2}{2\sigma_i^2}\right],
\end{equation}
with $C_1 \simeq 0.67, C_2 = 1 - C_1$ and $\sigma_1 \simeq 0.78, \sigma_2 = 1.80$. This suggests that as the number of legs of the ladder grows, the spin dynamics may cross over to more and more regular diffusion.

\begin{figure}[t!]
    \centering
    \includegraphics[width=1.0\columnwidth]{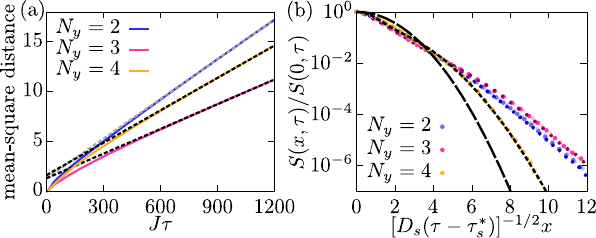}
    \vspace{-0.5cm}
    \caption{\textbf{Spin dynamics on wider ladders.} (a) Mean-square distance for indicated widths of $151 \times N_y$ cylindrical lattice [colored lines] with associated diffusive fits (dashed lines) for weak spin exchange, $\alpha = 0.08$. (b) Associated spin distribution vs rescaled spatial coordinate for indicated cylinder circumference $N_y$, revealing self-similar scaling. We have used data for $J\tau = 880, 1040, 1200$. In all cases, the distribution strongly deviates from Gaussian [$\sim \exp[-x^2/(4D_s\tau)]$ shown in long-dashed lines]. A sum of two Gaussian distributions [short dashed lines] describes the distribution well for $N_y = 4$. }
    \vspace{-0.5cm}
    \label{fig:multilegged_spin_ladder}
\end{figure}

We note, however, that such an extension of our results on two-leg ladders to two dimensions is seemingly in contradiction with recent results for a four-leg ladder in the isotropic Heisenberg limit ($\alpha = 1$) \cite{Guthardt2025}, in which a curious subdiffusive $\tau^{1/4}$ behavior is found at long times, instead of the more standard $\tau^{1/2}$ diffusive behavior that we presently find. While we presently do not know how to resolve this seeming discrepancy, we note that their MPS methodology is different from ours \footnote{There are two minor differences between the investigated models, both of which should not lead to qualitative differences. (1) Reference \cite{Guthardt2025} includes a nearest-neighbor density-density term which arises in the large $U/t$ limit of the Fermi-Hubbard model, which, however, leads to a fixed energy shift for a single dopant and thereby no differences in the dynamics. (2) The investigations in Ref.~\cite{Guthardt2025} is at lower spin couplings, $J/t < 1$.}. While we utilize a combination of METTS with TDVP and use TEBD for real-time evolution, they use a purification scheme \cite{Feiguin2005} along with TDVP for real-time evolution \cite{Haegeman2016,Haegeman2011}. In our studies, we have seen that we \emph{can} reproduce our TEBD results with TDVP but it requires that we take very special care of how the subspace expansion method \cite{Yang2020} is implemented. Additionally, we note that diffusive transport in the infinite temperature 2D Heisenberg model ($\alpha=1$) has been observed both experimentally in a cold atom quantum gas microscope experiment \cite{Wei_2022} and numerically in simulations on cylinders of width up to 8 \cite{anand2025}. The spin dynamics do not necessarily dictate the dopant dynamics, but at least in the case of the two-leg ladder studied here, we find qualitative agreement between them. Another possibility is that the fast delocalization associated with the lower spin couplings -- smaller $J/t$ -- investigated in Ref.~\cite{Guthardt2025} leads to more severe finite-size effects. 

\section{Conclusions} \label{sec.conclusions}
Using MPS calculations, we have thoroughly investigated the transport of a single charge (a hole) moving via nearest-neighbor hopping, $t$, in an XXZ spin ladder. We demonstrated that the charge is localized in the Ising limit, i.e. in the absence of spin-exchange processes, at all investigated nonzero temperatures and that the accompanying localization length scales with the underlying spin-spin correlation length. This greatly generalizes the results of the previously studied Hilbert space fragmented models \cite{Nielsen2023_3,Nielsen2024_1}. 

We have shown that spin-exchange processes at strength $\alpha$ lead to diffusive delocalization of charge and spin degrees of freedom alike. For both degrees of freedom, the associated diffusion coefficients $D_h, D_s$ ``turn on'' linearly in $\alpha$, i.e. $D_h, D_s \sim \alpha$ for $\alpha \ll 1$. For strong spin exchange $\alpha > 1$, we, however, find that the diffusion coefficient drops off as $\alpha^{-1}$, whereas the spins closely follow the behavior of the XX ladder with $D_s \simeq 0.95 \alpha J$ \cite{Steinigeweg2014,Kloss2018,Rakovszky2022}. The charge and spin distributions are, furthermore, demonstrated to follow self-similar scaling behaviors that strongly deviate from the Gaussian behavior of regular diffusion, especially for weak spin exchange, $\alpha \ll 1$. This suggests a simultaneous breakdown of simple hydrodynamics \cite{Kadanoff1963,Chaikin_Lubensky_1995} but with retained pre-thermal self-similar scaling \cite{Calabrese2005,Chandran2013,Maraga2015,Feldmeier2020,Joaquin2022,Berges2008,Bhattacharyya2020,Mikheev2023,Gazo2025}, which can potentially be explained by superstatistics with robust non-Gaussian behavior or diffusing diffusivity models with Gaussian behavior appearing on very long timescales \cite{Chubynsky2014,Chechkin2017,Metzler2020}. Moreover, we have shown robust diffusion over a wide range of high temperatures ($J/k_BT \lesssim 2$) and spin couplings ($J\gtrsim 2t$). 

Finally, we conjecture that these findings generalize to fully two-dimensional spin lattices. In particular, in the Ising limit we anticipate that the charge will undergo a very fast delocalization crossover as the para- to antiferromagnetic transition is crossed. Moreover, with spin exchange we find that the observed non-Gaussian diffusion is weakly dependent on the ladder width, and we see clues of a crossover to regular Gaussian diffusion as more legs are added to the system. 

Our findings show that Hilbert space fragmentation \cite{Moudgalya2022,Moudgalya2022b} in the present case is not a necessary condition to ensure localization, nor does the proximity to such a fragmented model necessarily lead to special behavior. Instead, we posit that it is the much more general appearance of an effective disordered potential in the Ising limit (in complete analogy to Anderson localization \cite{Anderson1958}), which holds the key to the localization effect, whereas strongly non-Gaussian diffusion appears in the \emph{proximity} of this localized phase. This exemplifies that while Hilbert space fragmentation is a valuable computational property for investigating novel non-equilibrium quantum many-body dynamics, the origin of the discovered effects in such a regime is not necessarily due to the fragmentation itself. 

In the future, we hope to explore how our present findings impacts the recently discovered spin disorder-induced pairing mechanism \cite{Nielsen2025_1} in the fragmented phase. Indeed, when spin exchange is weak, $\alpha J \ll t_\parallel$, we expect dopant pairing to be retained, because the internal dynamics of the pair $\sim t_\parallel$ is fast compared to the spin dynamics of the environment $\sim \alpha J$. For strong enough spin exchange, the dopants might unbind and give rise to non-trivial dependencies on the spin coupling. \\

\begin{acknowledgments}
The authors thank Steven R. White, J. Ignacio Cirac, Sarang Gopalakrishnan, Pavel Kos,  and Jens H. Nyhegn for stimulating discussions and important input. The numerical calculations are performed using the ITensor library in Julia \cite{Fishman2022} and the TeNPy library in Python \cite{Hauschild_2018, Hauschild_2024}. K.K.N. acknowledges support from the Carlsberg Foundation through a Carlsberg Reintegration Fellowship (Grant No. CF24-1214). S.A. is supported by the U.S. Department of Energy,
Office of Science, Office of Basic Energy Sciences, Materials Sciences and Engineering Division under Contract
No. DE-AC02-05-CH11231 (Theory of Materials program KC2301). Part of this research uses
the Lawrencium computational cluster provided by the Lawrence Berkeley National Laboratory (supported by the U.S. Department of Energy, Office of Basic Energy Sciences under Contract No. DE-AC02-05-CH11231). M.Y. is supported by the Distinguished Postdoc Fellowship of the Munich Center for Quantum Science and Technology (MCQST) funded by DFG under the Excellence Strategy EXC2111-390814868.
\end{acknowledgments} 

\emph{Data availability.} The data that support the findings of this article are openly available \cite{data_availability}.

\bibliography{ref_dopant_dynamics}

\appendix

\section{Fragmentation of \texorpdfstring{$t$-$J_z$}{t-Jz} models} \label{app:fragmentation}
In this Appendix, we construct the symmetry operators that correspond to the Hilbert space fragmentation of spin-$1/2$ lattices with Ising interactions and 1D hopping. Moreover, we compute the number of Krylov subspaces and show that this scales exponentially in system size. In particular, this applies to models of the form 
\begin{align} \label{eq:Ham_fragmentation}
\Ham =  &-\sum_{\braket{\bi,\bj}_x,\sigma} t_{\bi\bj} \hat{\cal{P}}\left[\hat{c}^\dagger_{\bi\sigma}\hat{c}_{\bj\sigma} + \hat{c}^\dagger_{\bj\sigma} \hat{c}_{\bi\sigma}\right]\hat{\cal{P}}\nn \\
&+ \sum_{\bi} J^{(1)}_\bi \hat{S}_\bi^{z} + \frac{1}{2}\sum_{\bi,\bj} J^{(2)}_{\bi\bj}\hat{S}_\bi^{z}\hat{S}_\bj^{z} + \dots,
\end{align} 
where the $\dots$ indicate arbitrary higher-order Ising terms. The crucial part here is that the hopping cannot lead to any \emph{loops} in the lattice and should only be to nearest neighbors. Here, we will take the particular case where the hopping is along one spatial direction, $x$, indicated by $\braket{\bi,\bj}_x$. The geometry of the lattice is otherwise completely general. Since there is no hopping across the legs of the lattice, $y$, it is clear that the number of spin-$\uparrow$ and -$\downarrow$ in each leg is an integral of motion. That is, $[\hat{N}_{y\sigma}, \Ham] = [\sum_x \hat{n}_{(x,y)\sigma}, \Ham] = 0$ for $\sigma = \uparrow,\downarrow$ and $\hat{n}_{(x,y)\sigma} = \hat{c}^\dagger_{(x,y)\sigma}\hat{c}_{(x,y)\sigma}$ the number of spin-$\sigma$ at site $(x,y)$. This gives in total $2\times N_y$ integrals of motion. \\

Moreover, the arguments given in Sec. \ref{sec.fragmentation_and_beyond} mean that the \emph{pattern} of the spins is unaltered in each leg of the lattice. To be more precise, we define the operators 
\begin{align}
\hat{S}_{1,y} &= \sum_{x_1=0}^{N_x-1} \!\!\hat{S}^z_{(x_1,y)} \!\!\prod_{x=0}^{x_1-1} \![1 \!-\! \hat{n}_{(x,y)}], \nn \\
\hat{S}_{2,y} &= \sum_{x_1=0}^{N_x-1} \!\hat{n}_{(x_1,y)} \!\!\!\prod_{x=0}^{x_1-1} \![1 \!-\! \hat{n}_{(x,y)}] \!\!\!\!\!\sum_{x_2 = x_1 + 1}^{N_x-1}\!\!\!\!\!\hat{S}^z_{(x_1,y)} \!\!\!\!\!\prod_{x=x_1+1}^{x_2-1}\!\!\![1 \!-\! \hat{n}_{(x,y)}],\nn \\
\hat{S}_{3,y} &= \dots
\end{align}
for each leg $y = 0,1,\dots,N_y - 1$ of the lattice. Here, the string $\prod_{x=0}^{x_1-1} [1 \!-\! \hat{n}_{(x,y)}]$ makes sure that there is only a nonzero contribution to the sum over $x_1$ at the position of the \emph{first spin}. For $x_1 = 0$ this is the empty product equal to $1$. Writing an Ising eigenstate as $\ket{\bsigma} = \prod_{\bi}\ket{\sigma_{\bi}}$, where $\sigma_\bi = +1/2,0,-1/2$ signifies a spin-$\uparrow$, a hole and a spin-$\downarrow$ respectively, it follows that
\begin{align}
\hat{S}_{l,y} \ket{\bsigma} &= \sigma_{x_l,y} \ket{\bsigma}.
\end{align}
Here, $x_l$ is the $x$-coordinate of the $l$th spin in leg $y$, i.e., the $l$th value of $\sigma_\bi$ that is nonzero. In total, we can, thus, define $N_x$ such operators. Moreover, because the Hamiltonian in Eq.~\eqref{eq:Ham_fragmentation} can at most move the spins along the $x$-direction and \emph{not alter their relative order}, it follows that 
\begin{align}
&\Ham \hat{S}_{l,y} \ket{\bsigma} = \sigma_{x_l,y} \Ham \ket{\bsigma} = \hat{S}_{l,y}\Ham \ket{\bsigma} \Rightarrow \nn \\
&[\Ham, \hat{S}_{l,y}] \ket{\bsigma} = 0.
\end{align}
Since the Ising eigenstates, $\ket{\bsigma}$, define a \emph{complete set of states}, it follows that $[\Ham, \hat{S}_{l,y}] = 0$. Since they also commute with $\hat{N}_{y\sigma}$, we can simultaneously diagonalize the Hamiltonian in terms of eigenstates of $\hat{N}_{y\sigma}$ and all the $\hat{S}_{l,y}$'s. Moreover, in a symmetry sector with $N_{y\uparrow}$ spin-$\uparrow$ and $N_{y\downarrow}$ spin-$\downarrow$ in leg $y$, there are $N_{y\uparrow} + N_{y\downarrow}$ nonzero eigenvalues $S_{l,y}$, $\{S_{1,y},S_{2,y}, \dots, S_{N_{y\uparrow} + N_{y\downarrow},y}, 0, \dots, 0\}$. Hence, for given $\{N_{y\uparrow},N_{y\downarrow}\}_{y=0}^{N_y-1}$, there are 
\begin{align}
\prod_{y = 0}^{N_y - 1} \binom{N_{y\uparrow} + N_{y\downarrow}}{N_{y\uparrow}}
\end{align}
symmetry sectors. The total number of Krylov subspaces is then
\begin{align}
K &= \sum_{\{N_{y\uparrow},N_{y\downarrow}\}_{y=0}^{N_y-1}}\prod_{y = 0}^{N_y - 1} \binom{N_{y\uparrow} + N_{y\downarrow}}{N_{y\uparrow}} \nn\\
&=\left[\sum_{\{N_{y\uparrow},N_{y\downarrow}\}}\!\!\!\!\binom{N_{y\uparrow} + N_{y\downarrow}}{N_{y\uparrow}} \right]^{N_y} \nn \\
&=\left[\sum_{N_{y\uparrow}=0}^{N_x}\sum_{N_{y\downarrow}=0}^{N_x-N_{y\uparrow}} \!\!\binom{N_{y\uparrow} + N_{y\downarrow}}{N_{y\uparrow}} \right]^{N_y} \nn \\
&=\left[\sum_{N_{y\uparrow}=0}^{N_x} \!\!\binom{N_x + 1}{N_{y\uparrow} + 1} \right]^{N_y} =\left[2^{N_x + 1} - 1 \right]^{N_y},
\end{align}
which is exponential in the total system size, $N = N_x \times N_y$. This establishes fragmentation \cite{Moudgalya2022b}. Moreover, we can check that this decomposition also gives the full Hilbert space dimension. Indeed, for each leg, $y$, with quantum numbers $N_{y\uparrow},N_{y\downarrow}$, the Krylov subspace has size
\begin{align}
D(N_{y\uparrow},N_{y\downarrow}) = D(N_{y\uparrow} + N_{y\downarrow}) = \binom{N_x}{N_{y\uparrow} + N_{y\downarrow}}.
\end{align}
The total Hilbert space dimensionality is then
\begin{align}
{\rm dim}({\cal H}) &= \!\!\sum_{\{N_{y\uparrow},N_{y\downarrow}\}_{y=0}^{N_y-1}}\!\prod_{y=0}^{N_y-1} \binom{N_x}{N_{y\uparrow} + N_{y\downarrow}} \binom{N_{y\uparrow} + N_{y\downarrow}}{N_{y\uparrow}} \nn \\
&= \prod_{y=0}^{N_y-1} \sum_{\{N_{y\uparrow},N_{y\downarrow}\}}\binom{N_x}{N_{y\uparrow} + N_{y\downarrow}} \binom{N_{y\uparrow} + N_{y\downarrow}}{N_{y\uparrow}} \nn \\
&= \left[\sum_{\{N_{y\uparrow},N_{y\downarrow}\}}\binom{N_x}{N_{y\uparrow} + N_{y\downarrow}} \binom{N_{y\uparrow} + N_{y\downarrow}}{N_{y\uparrow}}\right]^{N_y} \nn \\
&= \left[3^{N_x}\right]^{N_y} = 3^{N_x N_y}.
\end{align}
This is, indeed, the total size of the Hilbert space, as each site has a three-dimensional Hilbert space, $\ket{0},\ket{\uparrow},\ket{\downarrow}$. Finally, for a single hole, say in leg $y = 0$, $N_{y=0\uparrow} + N_{y=0\downarrow} = N_x - 1$, whereas  $N_{y\uparrow} + N_{y\downarrow} = N_x$. The total size of that Krylov subspace is, thus, 
\begin{align}
\!\!\!\prod_{y = 0}^{N_y-1} \!D(N_{y\uparrow} \!+\! N_{y\downarrow}) &= D(N_x \!-\! 1) = \binom{N_x}{N_x \!-\!1} = N_x.\!\!
\end{align}
This is, indeed, what is indicated in Fig.~\ref{fig:HSF}. The first step comes from the fact that $D(N_x) = 1$.

\section{Details on TEBD} \label{app:TEBD_details}
In this Appendix, we give more details on how the TEBD calculations are carried out.

\begin{figure}[ht!]
    \centering
    \includegraphics[width=1.0\columnwidth]{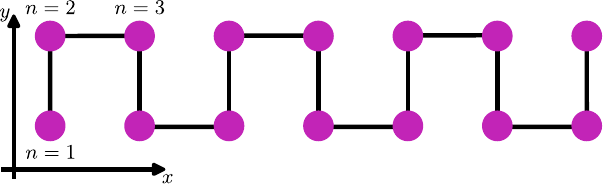}
    \vspace{-0.5cm}
    \caption{(a) Illustration of how the sites are connected into a snake pattern, mapping the two-leg ladder to a one-dimensional chain, $n = 1, 2, 3, \dots$}
    \vspace{-0.5cm}
    \label{fig:MPS_snake_pattern}
\end{figure}

To set up the TEBD algorithm, we follow Ref. \cite{White2015}. To this end, we write the Hamiltonian as the sum of a leg ($\parallel$) and rung ($\perp$)  Hamiltonian
\begin{align}
\Ham &= \Ham_\perp + \Ham_\parallel \nn \\
\Ham_\perp &= \sum_{x = 0}^{N_x-1} \Ham_{\perp,x}, \Ham_\parallel = \sum_{y=0}^1\sum_{x = 0}^{N_x-2} \Ham_{\parallel,x}^{(y)}
\end{align}
Here, $\Ham_{\perp,x}$ is the local term that connects the two sites $(x,0)$ and $(x,1)$ on rung $x$, whereas $\Ham_{\parallel,x}^{(y)}$ connects the two sites $(x,y),(x+1,y)$ on leg $y$. Moreover, we map the two-leg ladder with dimension $N_x \times 2$ to a one-dimensional line as indicated in Fig.~\ref{fig:MPS_snake_pattern}, typically referred to as snaking an MPS through the lattice. We then define the local Hamiltonian operators in this linear space ($n \in \{1,2, \dots, 2N_x-1\}$) as 
\begin{align} \label{eq:H_n}
\hat{H}_n = \left\{
\begin{matrix} 
\frac{1}{2}\Ham_{\perp,(n-1)/2}, &  n {\; \rm odd},  \\ 
\Ham_{\parallel,n/2}^{(1)}, &  n = 2 \mod 4, \\
\Ham_{\parallel,n/2}^{(0)}, &  n = 0 \mod 4.
\end{matrix}\right. 
\end{align}
Note that we use \emph{half} the rung coupling on the odd-numbered sites. The reason for this will be apparent shortly. Each time step of the TEBD is then performed by the following Trotterization. The first part consists of (step size $\delta\tau$) 
\begin{align}
\hat{U}(\delta \tau) = \prod_{n = 1}^{2N_x - 1} \exp\left[-i \hat{H}_n \delta \tau / 2\right].
\end{align}
In a linear chain, a TEBD time step then consists of \cite{White2004} updating the wave function at time $\tau$ to $\tau + \delta\tau$ with the update, $\ket{\psi(\tau + \delta \tau)} = \hat{U}(\delta \tau){\rm Rev}[\hat{U}(\delta \tau)] \ket{\psi(\tau)}$, where ${\rm Rev}[\cdot]$ reverses the order of the products. However, in the ladder we must take into account that some of the coupled sites are no longer along the line connecting the sites [Fig.~\ref{fig:MPS_snake_pattern}]. Therefore, the total TEBD time step consists of the following sequence of operations
\begin{align} \label{eq.TEBD_timestep}
\ket{\psi(\tau + \delta \tau)} {=}& \hat{U}(\delta \tau) \hat{S} \hat{U}(\delta \tau) \nn \\
& \times {\rm Rev}[\hat{U}(\delta \tau) \hat{S} \hat{U}(\delta \tau)]\ket{\psi(\tau)}· 
\end{align}
Here, the second part consists of ($t$-$J$ type) SWAP operations on all rungs (i.e. odd $n$)
\begin{align}
\hat{S} = \prod_{n {\; \rm odd}} \hat{S}_n.
\end{align}
The swapping of the fermions on the rungs, enable us to couple the sites along the ladder that were not coupled in the first operation of $\hat{U}(\delta\tau)$. Moreover, the second application of $\hat{U}(\delta \tau)$ after the SWAP explains why the Hamiltonian term for the rung part is cut in half in Eq.~\eqref{eq:H_n}. While all the Hamiltonian terms can straightforwardly be implemented in ITensor \cite{Fishman2022}, the $t$-$J$ type SWAP gates are not implemented beforehand. We do this by defining setting
\begin{align}
\bra{00} \hat{S}_n\ket{00} &= \bra{0\!\uparrow} \hat{S}_n\ket{0\!\uparrow} = \bra{0\!\downarrow} \hat{S}_n\ket{0\!\downarrow} = 1, \nn \\
\bra{\uparrow\uparrow} \hat{S}_n\ket{\uparrow\uparrow} &= \bra{\downarrow\uparrow} \hat{S}_n\ket{\downarrow\uparrow} = \bra{\downarrow\downarrow} \hat{S}_n\ket{\downarrow\downarrow} = -1, 
\end{align}
and using that $\hat{S}_n^\dagger = \hat{S}_n$. All other matrix elements are $0$. This is then implemented as an ITensor operator of SiteType "tJ".

The error in each Trotter step in Eq. \eqref{eq.TEBD_timestep} is $\calO[(\delta \tau)^3]$ such that the overall error per unit time is $\calO[(\delta \tau)^2]$ \cite{White2015,Stoudenmire2010}. We always use time steps of $t_\parallel\delta \tau \leq 0.1$. For large $\alpha J$, we reduce the timestep further to ensure that $\alpha J / 4 \times \delta \tau \leq 0.125$. In this manner, the Trotter error even in the spin exchange terms is negligible. Indeed, we have checked the dependency on time step size in several instances and found no discernible differences.

\section{Subdiffusive vs diffusive fits} \label{app:subdiffusive_fits}
{In this Appendix, we give an explicit example of how fitting a na{\"i}ve power-law leads to subdiffusive behavior. We also explain why the data are better explained by (non-Gaussian) diffusion.

{In Fig.~\ref{fig:subdiff_vs_diff}(a), we repeat the data from Fig.~\ref{fig:zero_vs_inf_temp_alpha}(a) in the cases of $\alpha = 0.08,0.3$. We show the subdiffusive fits obtained by fitting a straight line to the log-transformed distance vs the log-transformed time:  $\ln[d_{\rm rms}(\tau)]$ vs $\ln(t_\parallel\tau)$. We also show the diffusive fits used in the main text coming from fitting a straight line to the mean-squared distance vs time -- $[d_{\rm rms}(\tau)]^2$ vs $t_\parallel\tau$. From these fits alone, it is not obvious which one describes the data the best, although the diffusive fit for $\alpha = 0.08$ seems to follow the data better down to shorter times than the subdiffusive fit.

{However, we expect the hole density distribution to follow a scaling behavior in the diffusive -- or subdiffusive -- regime. We check this in Figs.~\ref{fig:zero_vs_inf_temp_alpha}(b) and \ref{fig:zero_vs_inf_temp_alpha}(c). Importantly, the diffusive fits show a considerably better data collapse than the subdiffusive fits. This shows that the time offsets $\tau^*_h$ are necessary to get a proper collapse of the data. Importantly, with such $\tau^*_h$, Fig.~\ref{fig:zero_vs_inf_temp_alpha}(a) shows that the simpler diffusive fit works just as well -- if not slightly better -- than the subdiffusive fits. We see similar behavior over the entire investigated range of spin exchanges, $\alpha$. Therefore, the best and simplest explanation of the data is that the hole \emph{diffuses} for any nonzero spin exchange, $\alpha$. A similar analysis holds for spin transport, for which we find qualitatively the same behavior.

\begin{figure}[ht!]
    \centering
    \includegraphics[width=1.0\columnwidth]{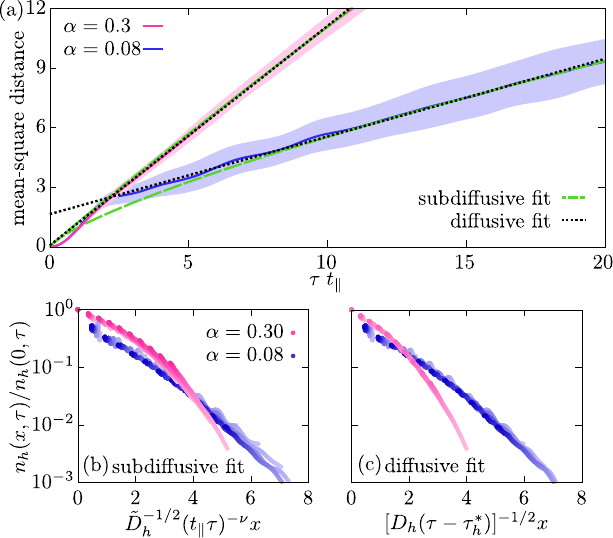}
    \vspace{-0.5cm}
    \caption{\textbf{Subdiffusive vs diffusive fits.} (a) Mean-square distance for indicated values of $\alpha$ at $J=5t_\parallel$ and $t_\perp = 0$ repeated from Fig.~\ref{fig:zero_vs_inf_temp_alpha}(b). Green long-dashed lines show the subdiffusive fits $[d_{\rm rms}(\tau)]^2 = 2 \tilde{D}_h (t_\parallel \tau)^{2\nu}$, with $\nu(\alpha = 0.08) \simeq 0.38$ and $\nu(\alpha = 0.3) \simeq 0.48$. Black short-dashed lines show the diffusive fits $[d_{\rm rms}(\tau)]^2 = 2 D_h (\tau - \tau^*_h)$. (b) Scaled hole density distribution vs the scaled coordinate $\tilde{D}_h^{-1/2} (t_\parallel\tau)^{-\nu}x$ coming from the subdiffusive fits. (c) Scaled hole density distribution vs the scaled coordinate $[D_h (\tau - \tau^*_h)]^{-1/2}x$ coming from the diffusive fits. The color gradient on the data points denote the evolution time from shortest used (bright colors) to longest used (dark colors). We use time intervals of $t_\parallel\tau \in [5,20]$, $t_\parallel\tau \in [5,15]$ for $\alpha = 0.08$ and $\alpha = 0.3$, respectively.}
    \vspace{-0.5cm}
    \label{fig:subdiff_vs_diff}
\end{figure}

\section{Details on METTS} \label{app:METTS_details}
In this Appendix, we give more details on how the METTS sampling of the spin background is performed, as well as an analysis of the autocorrelation time for the achieved samples. 

For our METTS algorithm to sample the spin background, we use an imaginary time step of $\delta\beta = 0.05/J$. The algorithm starts from the N{\'e}el-ordered state 
\begin{align} \label{eq.Neel_ordered_state}
\ket{\bsigma_0} = \ket{\substack{
\downarrow\uparrow\downarrow\uparrow\downarrow\uparrow\downarrow\uparrow\\
\uparrow\downarrow\uparrow\downarrow\uparrow\downarrow\uparrow\downarrow 
}\dots}.
\end{align}
The achieved METTS at each step, $j$, $\ket{\phi(\bsigma_j)} \propto \exp[-\beta \Ham_J/2]\ket{\bsigma_j}$ is collapsed in the $\hat{S}^{z}$ eigen-basis for odd $j$, whereas it is collapsed in the orthogonal $\hat{S}^{x}$ eigen-basis for even $j$. This is to minimize the sample-to-sample autocorrelation \cite{Stoudenmire2010,Wietek2021} and also results in fluctuations in the $S^z$ charge (the magnetization) of the sampled spin state. Moreover, we let the system thermalize by using $10$ "warm-up" steps, which are not saved as samples. Finally, to minimize autocorrelation even further, we only keep every fourth sample, all of which are collapsed in the $\hat{S}^{z}$ eigen-basis. In Fig.~\ref{fig:autocorrelation_function}, we show the estimated autocorrelation function \cite{Wietek2021}
\begin{align} \label{eq:autocorrelation_function}
\rho[\hat{A}](l) = \frac{1}{|S| - l} \sum_{j = 1}^{|S| - l}(A_j - \mathbb{E}[A])(A_{j+l} - \mathbb{E}[A]),
\end{align}
based on this sampling for the average energy per site, $\hat{A} = \Ham_J / N$, for the used $|S| = 100$ samples at two indicated inverse temperatures. Here, $A_j = \bra{\phi_j} \hat{A}\ket{\phi_j}$ is the average of $\hat{A}$ for step $j$, whereas $\mathbb{E}[A] = |S|^{-1} \sum_j^{|S|} A_j$ is the sample mean. We find no discernible correlations between the samples. 

\begin{figure}[h!]
    \centering
    \includegraphics[width=1.0\columnwidth]{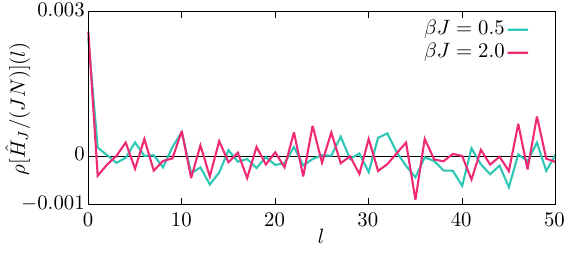}
    \vspace{-0.5cm}
    \caption{Autocorrelation function [Eq.~\eqref{eq:autocorrelation_function}] for indicated inverse temperatures at $\alpha = 0.08$.}
    \vspace{-0.5cm}
    \label{fig:autocorrelation_function}
\end{figure}

{In total then, the combined METTS sampling + TEBD time evolution proceeds as follows:
\begin{itemize}
    \item Using TDVP and the above-described METTS algorithm, we obtain 100 spin samples, $\ket{\bsigma}$, all projected in the $S^z$ eigen-basis. 
    \item For each of these, we use TEBD, as described in Appendix \ref{app:TEBD_details}, to first do the imaginary time evolution to obtain the METTS: $\ket{\phi(\bsigma)} = p^{-1/2}(\bsigma) \exp[-\beta \Ham_J/2]\ket{\bsigma_j}$, with $p(\bsigma) = \bra{\bsigma} \exp[-\beta \Ham_J] \ket{\bsigma}$. 
    \item We then remove a single spin, $\sigma_{\bf 0}$, from $\bf{i} = \bf{0}$ and use TEBD for real-time evolution, as described in Appendix \ref{app:TEBD_details}. In this manner, we obtain the time-evolved sample state $\ket{\phi_{\sigma_{\bf 0}}(\bsigma,\tau)} = e^{-i\Ham\tau} \hat{c}_{{\bf 0},\sigma_{\bf 0}}\ket{\phi(\bsigma)}$ in Eq.~\eqref{eq.sampled_pure_state}.
\end{itemize}
The reason why we split the METTS sampling from the time evolution is that it is the latter process that is the bottleneck of the computation. Indeed the entire METTS sampling takes a few hours, whereas the time evolution of \emph{each sample} takes several days. By splitting the two, we can, hereby, parallelize the time evolution over the 100 created samples.

\section{Magnetic energy fluctuations} \label{app:magnetic_energy_fluctuations}
In this Appendix, we calculate the magnetic energy fluctuations in a single dopant hopping event to qualitatively describe the spin-exchange dependency of the activation energy appearing in the Arrhenius relation for the diffusion coefficient. 

We start from a state with a single hole at site $\bi = (x,0)$
\begin{align}
\hat{\rho}_\bi = \sum_{\sigma} \hat{c}_{\bi\sigma}\hat{\rho}_J\hat{c}^\dagger_{\bi\sigma},
\end{align}
where $\rho_J = \exp[-\beta \Ham_J]/{\rm tr}(\exp[-\beta \Ham_J])$ is the spin Gibbs state at inverse temperature $\beta$. Note that the state with a single dopant at the neighboring site $\bi + \hat{x} = (x+1,0)$ can be achieved in the following manner
\begin{align}
\hat{\rho}_{\bi + \hat{x}} = \sum_{\sigma} \hat{c}_{\bi + \hat{x}\sigma}\hat{\rho}_J\hat{c}^\dagger_{\bi + \hat{x}\sigma} = \hat{V}^\dagger \hat{\rho}_\bi \hat{V},
\end{align}
with $\hat{V} = \hat{\calP} \sum_\sigma \hat{c}^\dagger_{\bi + \hat{x}\sigma}\hat{c}_{\bi\sigma} \hat{\calP}$ (remember that $\hat{\calP}$ is the projection operator onto the subspace with at most one fermion per site). This means that the mean change in magnetic energy in a hopping event can be expressed as
\begin{align} \label{eq:change_mean}
\mu(\Delta E_J) = {\rm tr}(\hat{\rho}_\bj \hat{H}_J) - {\rm tr}(\hat{\rho}_\bi \hat{H}_J) = {\rm tr}(\hat{\rho}_\bi \Delta \Ham_J),
\end{align}
with $\Delta \Ham_J = \hat{V} \Ham_J \hat{V}^\dagger - \Ham_J$. One may check that ${\rm tr}(\hat{\rho}_\bj \hat{H}_J) = {\rm tr}(\hat{\rho}_\bi \hat{H}_J) = 0$ at $\beta = 0$, as the expressions depend on spin-spin correlators which vanish at infinite temperatures. Hence, the mean change in energy is also $0$. Moreover, the rewriting in Eq.~\eqref{eq:change_mean} allows us to think of the hopping event as a transformation on the Hamiltonian, rather than on the state. It also means that we can straightforwardly express the \emph{fluctuations} in the energy change as the variance of $\Delta \Ham_J$
\begin{align} \label{eq:change_variance}
{\rm Var}(\Delta E_J) = {\rm Var}(\Delta \Ham_J) = {\rm tr}\big(\hat{\rho}_\bi [\Delta \Ham_J - \mu(\Delta E_J)]^2\big).
\end{align}
Since $\hat{V}$ moves a fermion from site $\bi$ to site $\bi + \hat{x}$ (and $\hat{V}^\dagger$ vice versa), it follows that $\hat{V}\hat{S}^{l}_\bi\hat{V}^\dagger = \hat{S}^l_{\bi + \hat{x}}$ and $\hat{V}\hat{S}^{l}_{\bi+\hat{x}}\hat{V}^\dagger = 0$ and leaves all other couplings unaltered. As a result,
\begin{align} \label{eq:Hamiltonian_change}
\Delta \Ham_J ={}& \Ham_J(\bi - \hat{x},\bi + \hat{x}) + \Ham_J(\bi + \hat{y},\bi + \hat{x}) \nn \\
&- \sum_{\bj \in {\rm NN}(\bi + \hat{x})} \!\!\!\!\Ham_J(\bj,\bi + \hat{x}),
\end{align}
where $\Ham_J(\bi,\bj) = J[S^z_\bi S^z_\bj + \alpha ( S^x_\bi S^x_\bj  + S^y_\bi S^y_\bj) ]$, and ${\rm NN}(\bj)$ are the set of sites that are nearest neighbors to $\bj$. At infinite temperatures (where $\mu(\Delta E_J) = 0$), we may then write Eq.~\eqref{eq:change_variance} as
\begin{align} \label{eq:change_variance_2}
{\rm Var}(\Delta E_J) ={} &{\rm tr}\big( \hat{\rho}_\bi [\Ham_J(\bi - \hat{x},\bi + \hat{x})]^2 \big) \nn \\
+ &{\rm tr}\big( \hat{\rho}_\bi [\Ham_J(\bi + \hat{y},\bi + \hat{x})]^2 \big)\nn \\
+& \sum_{\bj\in{\rm NN}(\bi + \hat{x})}{\rm tr}\big( \hat{\rho}_\bi [\Ham_J(\bj,\bi + \hat{x})]^2 \big).
\end{align}
Here, we use that only the quadratic terms give a nonzero contribution, because the rest will depend on spin-spin correlator that are zero at infinite temperature. This also means that four of the five terms above (there are three terms in the sum over $\bj$) give exact the same contribution. The fifth and last one depends on $\Ham_J(\bi,\bi+\hat{x})$ and does not contribute, because there is a hole at site $\bi$ in the state $\hat{\rho}_\bi$. Finally, it is only terms that are quadratic in the spin operators on each site. Therefore,
\begin{align} \label{eq:change_variance_3}
&{\rm Var}(\Delta E_J) = 4 {\rm tr}\big( \hat{\rho}_\bi [\Ham_J(\bi - \hat{x},\bi + \hat{x})]^2 \big) \nn \\
&= 4J^2 \bigg( {\rm tr}\big(\hat{\rho}_\bi[\hat{S}^z_{\bi - \hat{x}}\hat{S}^z_{\bi + \hat{x}}]^2\big) \nn \\
&+ \alpha^2\Big[{\rm tr}\big(\hat{\rho}_\bi[\hat{S}^x_{\bi - \hat{x}}\hat{S}^x_{\bi + \hat{x}}]^2\big) + {\rm tr}\big(\hat{\rho}_\bi[\hat{S}^y_{\bi - \hat{x}}\hat{S}^y_{\bi + \hat{x}}]^2\big)\Big] \bigg) \nn \\
&= 4J^2 \bigg( \frac{1}{16} + \alpha^2\Big[\frac{1}{16} +\frac{1}{16}\Big] \bigg) = \frac{J^2}{4}\left[1 + 2\alpha^2\right].
\end{align}
Taking the square root yields the standard deviation ${\rm std}[\Delta E_J] = J[1 + 2\alpha^2]^{1/2}/2$ of the magnetic energy fluctuations, which is used in Sec. \ref{sec.spin_exchange_temperature_dependency}.

\end{document}